\documentclass[ superscriptaddress, aps, prl, twocolumn, floatfix ]{revtex4-2}

\usepackage{amsmath, amsfonts, amssymb, amsthm, amstext, bm, bbm}
\usepackage[colorlinks=true, linkcolor=blue,citecolor=red, urlcolor=blue, hypertexnames=true]{hyperref}
\usepackage{graphicx, subcaption}

\usepackage{ragged2e}
\DeclareCaptionJustification{justified}{\justifying}
\newcommand{\brac}[1]{\left( #1 \right) }

\begin{document}

\title{Robustness of Kardar-Parisi-Zhang scaling in a classical integrable spin chain with broken integrability}

\author{Dipankar Roy}
\email{dipankar.roy@icts.res.in}
\affiliation{International Centre for Theoretical Sciences, Tata Institute of Fundamental Research, Bangalore 560089, India}
\author{Abhishek Dhar}
\email{abhishek.dhar@icts.res.in}
\affiliation{International Centre for Theoretical Sciences, Tata Institute of Fundamental Research, Bangalore 560089, India}
\author{Herbert Spohn}
\email{spohn@ma.tum.de}
\affiliation{Zentrum Mathematik and Physik Department, Technische Universität München, Garching 85748, Germany}
\author{Manas Kulkarni}
\email{manas.kulkarni@icts.res.in}
\affiliation{International Centre for Theoretical Sciences, Tata Institute of Fundamental Research, Bangalore 560089, India}

\date{\today}

\begin{abstract}
	Recent investigations have observed superdiffusion in integrable classical and quantum spin chains. An intriguing connection between these spin chains and Kardar-Parisi-Zhang (KPZ) universality class has emerged. Theoretical developments (e.g. generalized hydrodynamics) have highlighted the role of integrability as well as spin-symmetry in KPZ behaviour. However understanding their precise role on superdiffusive transport still remains a challenging task. The widely used quantum spin chain platform comes with severe numerical limitations. To circumvent this barrier, we focus on a classical integrable spin chain which was shown to have deep analogy with the quantum spin-$\frac{1}{2}$ Heisenberg chain. Remarkably, we find that KPZ behaviour prevails even when one considers integrability-breaking but spin-symmetry preserving terms, strongly indicating that spin-symmetry plays a central role even in the non-perturbative regime. On the other hand, in the non-perturbative regime, we find that energy correlations exhibit clear diffusive behaviour. We also study the classical analog of out-of-time-ordered correlator (OTOC) and  Lyapunov exponents. We find significant presence of chaos for the integrability-broken cases even though KPZ behaviour remains robust. The robustness of KPZ behaviour is demonstrated for a wide class of spin-symmetry preserving integrability-breaking terms.
\end{abstract}

\maketitle

Superdiffusive spin dynamics in 1D spin chains has garnered a lot of attention recently. In particular, anomalous spin transport has been observed in an \emph{integrable} model, namely the quantum Heisenberg spin-$\tfrac{1}{2}$ chain with \emph{isotropic} interactions at infinite temperature~\cite{2011-znidaric, 2017-ljubotina--prosen}. Subsequent numerical computations~\cite{2019-ljubotina--prosen} have shown that the spin correlation agrees with the exact correlation function~\cite{2004-prahofer--spohn} known in the context of the 1D Kardar-Parisi-Zhang (KPZ) universality class~\cite{1986-kardar--zhang, 2018-takeuchi}. Similar properties have been unearthed in an integrable quantum spin chain with larger symmetry group~\cite{2020-dupont-moore}. This connection between integrability and KPZ superdiffusion has also been a topic of recent analytical studies in the context of quantum models~\cite{2016-alvaredo--yoshimura, 2018-ilievski--prosen, 2019-gopalakrishnan-vasseur, 2020-nardis--vasseur, 2021-ilievski--ware, 2021-bulchandani--ilievski}. Interestingly, recent experimental results have provided the evidence of 1D KPZ physics in quantum spin chains as well~\cite{2021-scheie--tennant, 2021-wei--zeiher}. Moreover, numerical studies have also revealed similar characteristics for the spin transport and correlations in integrable and isotropic \emph{classical} models~\cite{2013-prosen-zunkovic, 2019-das--dhar, 2020-krajnik-prosen, 2020-krajnik--prosen}. These developments in 1D quantum and classical spin chains suggest that both spin symmetry and integrability have pivotal implications on the existence and nature of superdiffusion. It has been argued~\cite{2020-nardis--vasseur} that the quantum-classical correspondence is related to the dominant role of solitons (analogous to string excitations in quantum case) in causing superdiffusive behaviour.

Naturally allied to the integrability property in the 1D spin chains is the question: \emph{What happens to the KPZ superdiffusion when integrability is broken?} Perturbation theory has been applied for understanding the fate of superdiffusion in the quantum Heisenberg spin-$\tfrac{1}{2}$ chain under the effect of weak integrability breaking perturbations~\cite{2021-nardis--ware}. On the other hand, in the strongly chaotic regime,
regular diffusion has been observed for spin transport at infinite temperature by using conventional perturbative methods~\cite{2021-claeys--arbeitman} where the integrable term was treated as perturbation. An extensive study of this problem is, nevertheless, still lacking in the literature. In particular, only perturbative regimes (where the weak parameter is either the integrability-breaking term or the integrable term itself) have been investigated and non-perturbative regimes are far from being understood. Needless to mention, quantum models are plagued by severe numerical limitations for such studies thereby motivating the use of classical integrable systems (which share properties analogous to quantum chains) as one of the most promising alternative platform.

In this Letter, we report that the KPZ superdiffusion is \emph{robust} when a \emph{symmetry preserving} interaction breaks integrability. This holds true even when integrability is broken strongly (non-perturbative).
Our assertion is based on an extensive numerical study for a 1D classical spin chain which involves the Hamiltonian of the \emph{integrable lattice Landau-Lifshitz} (ILLL) model~\cite{1982-sklyanin, 1988-sklyanin, 2007-faddeev-takhtajan} at the isotropic point and a spin-symmetry preserving, but integrability-breaking interaction (described in details later). When the integrability breaking term does not respect spin-symmetry we find significant deviations from KPZ behaviour~\cite{2022-supp}. We consider both perturbative and non-perturbative regimes. Since there exists a strong evidence of a classical-quantum correspondence~\cite{2019-das--dhar, 2020-nardis--vasseur}, we expect a similar behaviour in the quantum case. Our numerical simulations in the classical case allow us to probe transport and correlations for spin and energy for perturbations of different magnitude and different kinds of interactions. We find that the energy correlations exhibit  ballistic or diffusive behaviour depending on the strength of the perturbation.

\begin{figure}[htbp!]
			\begin{subfigure}{1\linewidth}
				\includegraphics[width=1\linewidth]{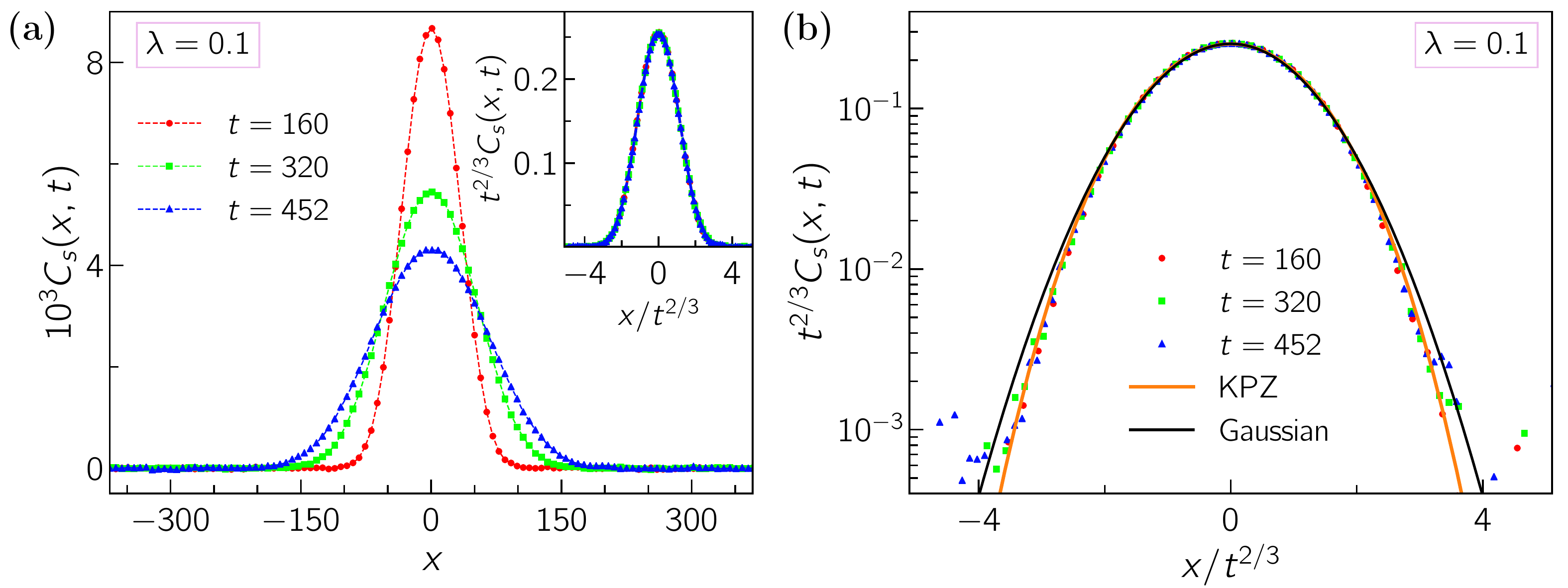}
			\end{subfigure}
			\begin{subfigure}{1\linewidth}
				\includegraphics[width=1\linewidth]{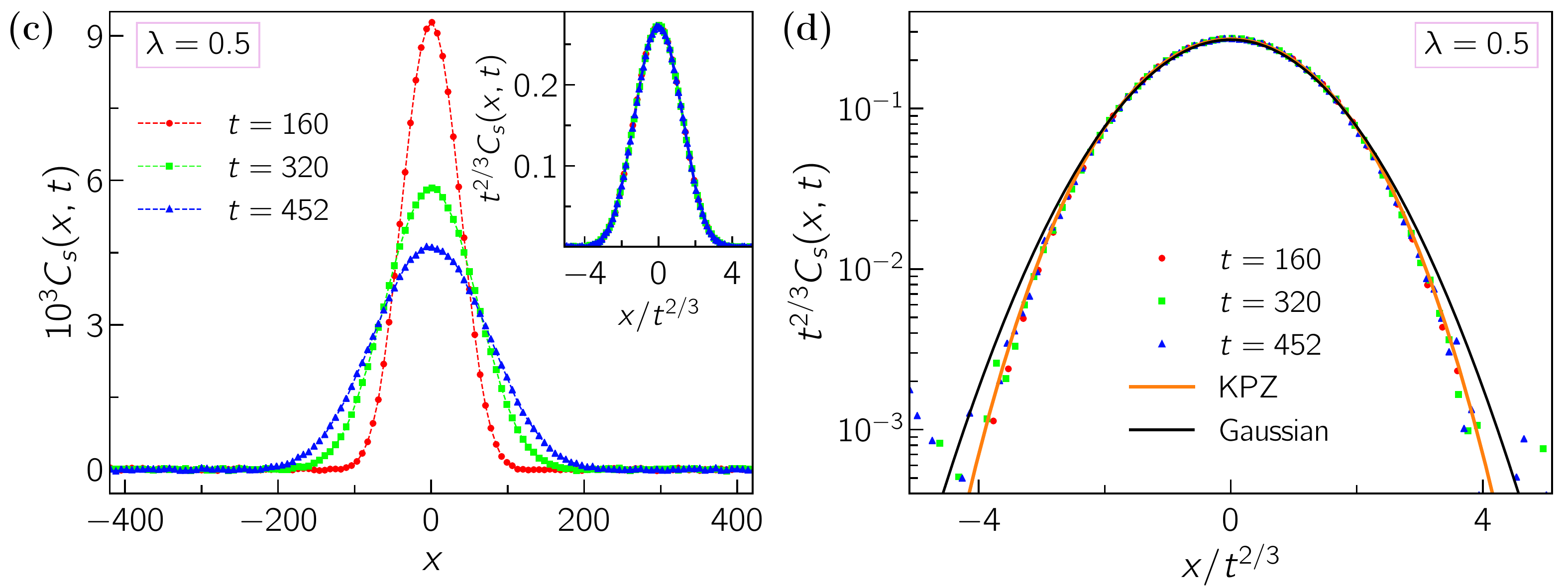}
			\end{subfigure}
			\begin{subfigure}{1\linewidth}
				\includegraphics[width=1\linewidth]{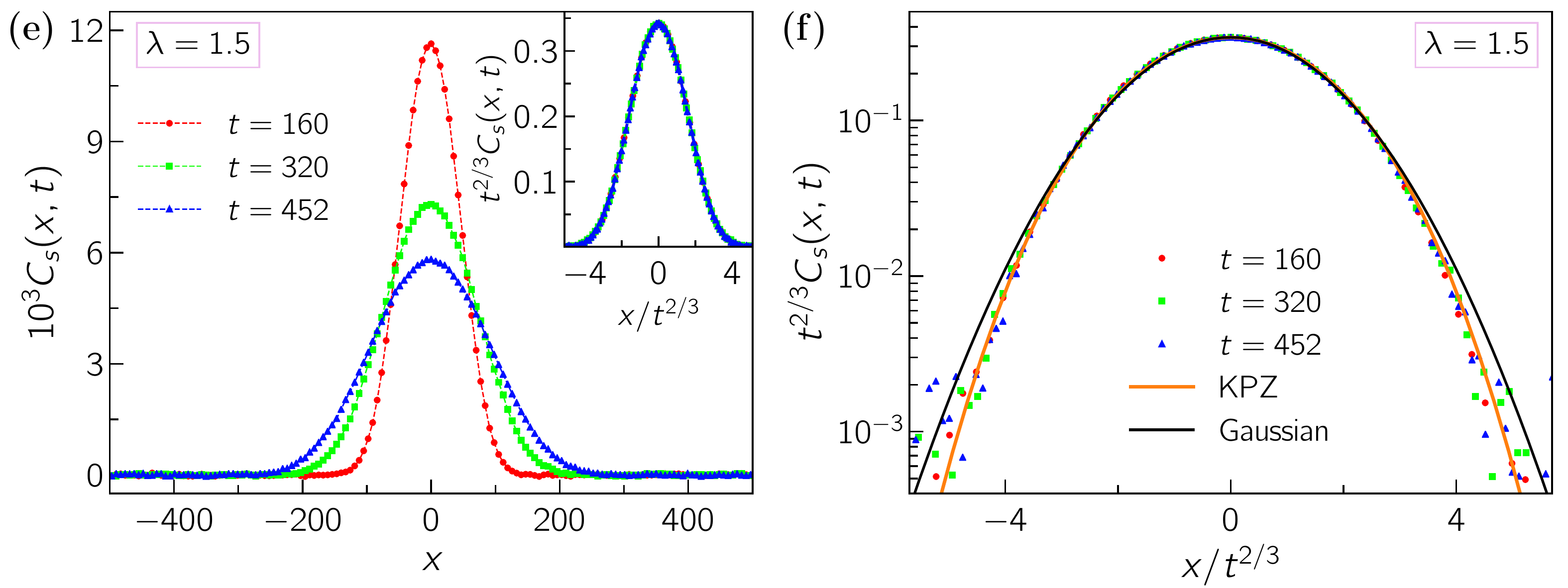}
			\end{subfigure}%
			\caption{(Color online) Plots of the spin correlations for different values of $\lambda$. We plot $C_{s}(x,t)$ versus $x$ in (a), (c), and (e) 
			and $t^{2/3}C_{s}(x,t)$ versus $x/t^{2/3}$ in (b), (d), and (f). We also plot the exact KPZ correlation function \cite{2004-prahofer--spohn} as well as a Gaussian function in (b), (d), and (f) for a comparison.
			Insets in (a), (c), and (e), show the collapse using KPZ exponent on a normal scale. 
			Total number of independent realizations is $2\times 10^5$ and $N=2048$.}
			\label{fig:spin}
\end{figure}

We consider a one-dimensional periodic chain of three-component classical spins $\vec{S}$ of unit length. The Hamiltonian is given by
\begin{equation}
	H 
	= 
	- \sum_{n=1}^{N} \brac{ J \ln \brac{ 1 + \vec{S}_{n} \cdot  \vec{S}_{n+1} } +  \lambda \vec{S}_{n} \cdot  \vec{S}_{n+1} },
	\label{eq:ham}
\end{equation} 
where $N$ is the length of the spin chain, $J$ is the strength of the integrable part, and $\lambda$ is the strength of the integrability-breaking perturbation. We set $N=2048$ and $J=1$ in our computations unless otherwise mentioned. The ILLL spin chain at the isotropic point, where the KPZ phenomenology has been observed recently~\cite{2019-das--dhar}, is recovered for $\lambda=0$. Thus we refer to our model described by the Hamiltonian in Eq.~\eqref{eq:ham} as the \emph{isotropic perturbed} ILLL (\textit{ip}ILLL) model. Notice that the Hamiltonian in Eq.~\eqref{eq:ham} remains invariant under a global rotation of spin vectors thereby obeying spin rotation symmetry. The spin dynamics in this system is determined by Hamilton's equations of motion
\begin{equation}
	\frac{ \text{d} \vec{S}_n }{ \text{d}  t } 
	= 
	\left\lbrace  \vec{S}_n , H \right\rbrace =  \vec{S}_n \times \vec{B}_n , \ \vec{B}_n = - \vec{ \nabla }_{ \vec{S}_n } H .
	\label{eq:dyn}
\end{equation}
In order to understand transport properties for a conserved quantity $q=\sum_{n=1}^{N}q_n$, we compute $C_{q}(x,t)$, the connected correlator for $q$, defined as 
\begin{equation}
\begin{aligned}
	C_{q}(x,t) 
	= \langle \left[ q_{x}(t) - \langle q_{0}(0) \rangle_{eq} \right] \left[ q_{0}(0) - \langle q_{0}(0) \rangle_{eq} \right]   \rangle_{eq} .
\end{aligned}
\end{equation} 
Here $\langle \cdot \rangle_{eq}$ denotes average with respect to the equilibrium distribution $e^{-\beta H}/Z$, where $Z$ is the partition function at temperature $T$ and $\beta=1/T$ is the inverse temperature. We are interested in the spin correlation $C_s(x,t)$ for $S^{z}$, the $z$-component of the spin $\vec{S}$, and the energy correlation $C_e(x,t)$ associated with the local energy defined as
\begin{equation}
	e_n = - J \ln \left( 1 + \vec{S}_{n} \cdot  \vec{S}_{n+1} \right) - \lambda \vec{S}_n \cdot \vec{S}_{n+1} .
\end{equation}
We expect that the correlation $C_q(x,t)$ satisfies
\begin{equation}
	C_{q}(x,t) = \frac{1}{ t^{\alpha} } \, f^q \! \Big(  \tfrac{x-ct}{t^{\alpha}} \Big) , 
\end{equation}
where $f^q \! \left(  \cdot \right)$ is a scaling function %(which may in general depend on the parameters $J, \lambda,$ and $\beta$) 
and $\alpha>0$ the scaling exponent. It is worth noting that unlike in nonlinear fluctuating hydrodynamics description for generic non-integrable models, where KPZ behaviour is associated with sound modes~\cite{2020-das--spohn}, here we have $c=0$. The exponent $\alpha$ can be directly extracted from the mean squared deviation (MSD) for $q$
\begin{equation}
	\langle \Delta x ^2 \rangle_q  := \sum_{x=1}^{N} x^2 C_{q}(x,t) \ \propto t^{ 2 \alpha} .
\end{equation}
To evaluate numerically these quantities (energy and spin correlations as well as corresponding MSDs) for the \textit{ip}ILLL spin chain, we perform numerical simulations that evolve the spin chain starting from equilibrium initial conditions at the chosen temperature. We then average over these equilibrium initial conditions to obtain our results. See supplementary material~\cite{2022-supp} for more details regarding the simulation methods.

\begin{figure}[htbp!]
	\begin{center}
		\begin{subfigure}{1\linewidth}
			\includegraphics[width=1\linewidth]{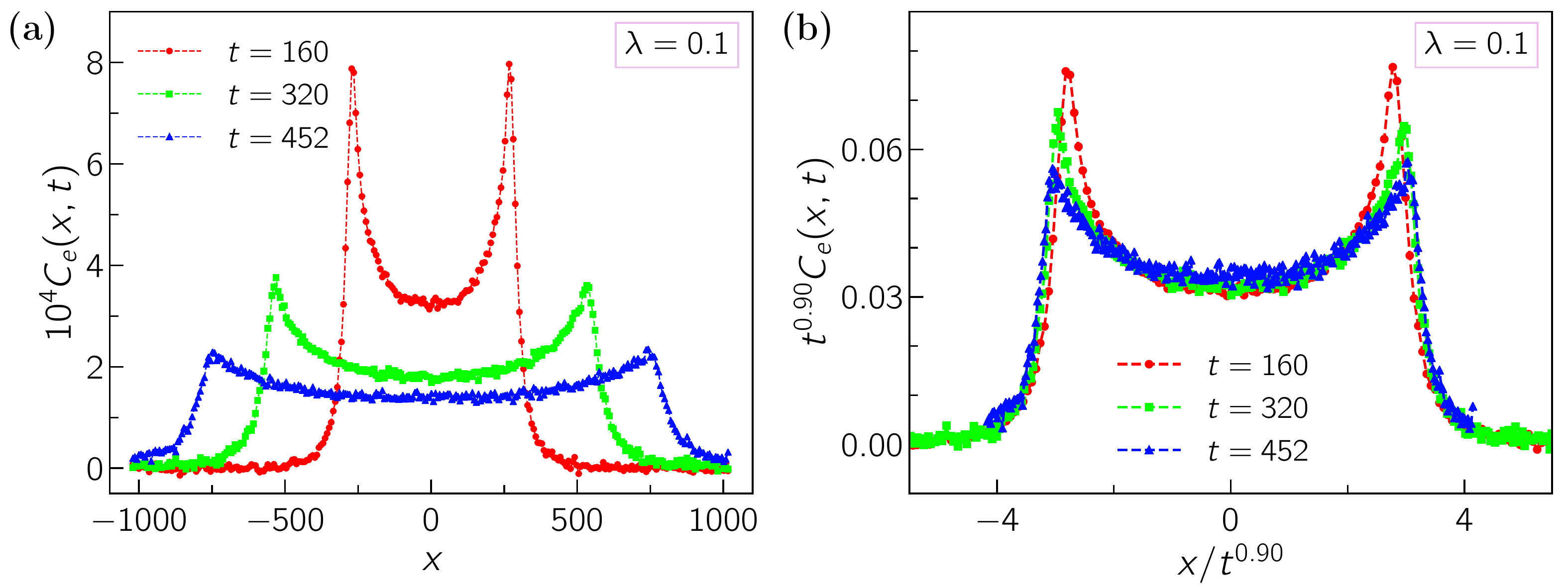}
		\end{subfigure}
		\begin{subfigure}{1\linewidth}
			\includegraphics[width=1\linewidth]{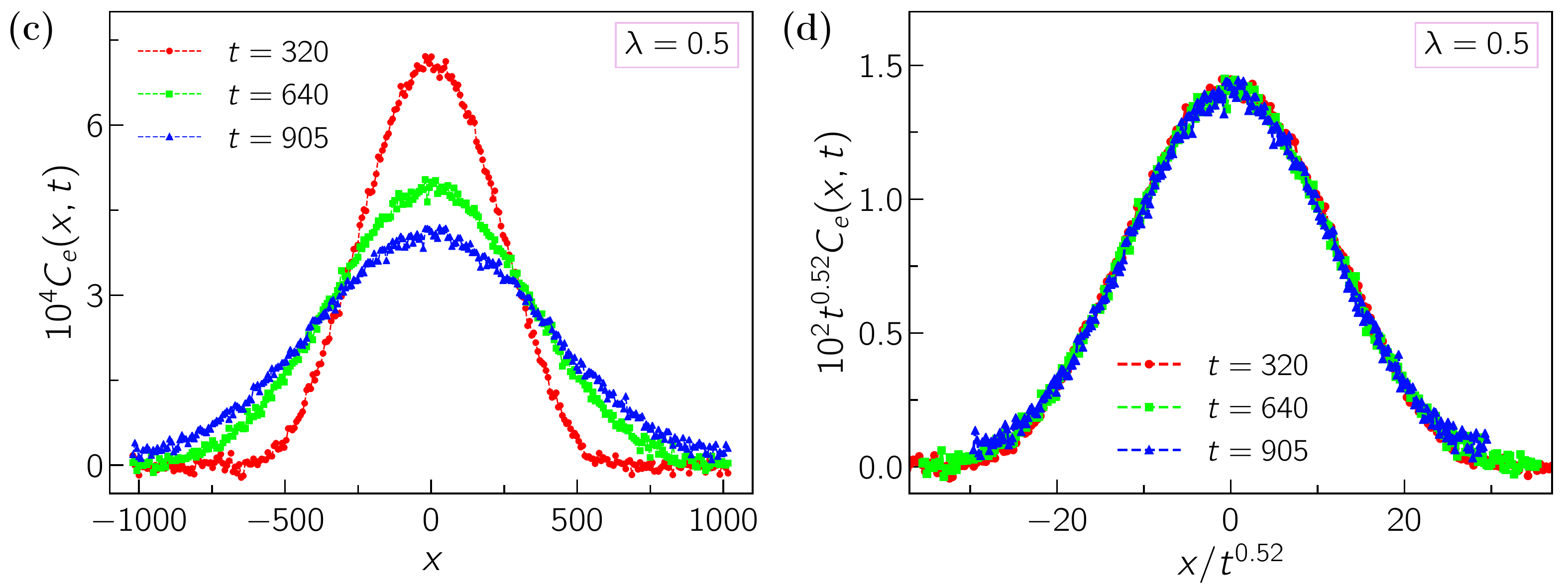}
		\end{subfigure}
		\begin{subfigure}{1\linewidth}
			\includegraphics[width=1\linewidth]{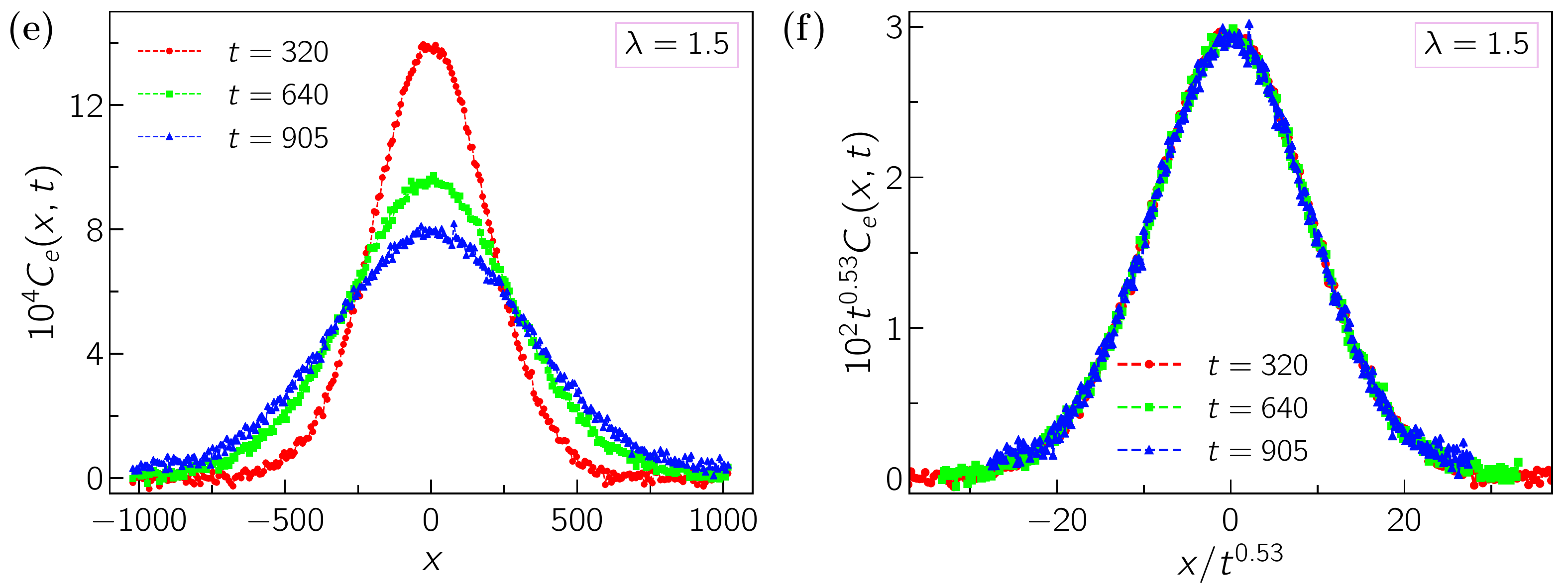}
		\end{subfigure}
	\end{center}
	\caption{(Color online) Plots of the energy correlations for different values of $\lambda$. We plot $C_{e}(x,t)$ versus $x$ in (a), (c), and (e). In (b), we plot $t^{0.9}C_{e}(x,t)$ versus $x/t^{0.9}$, in (d), we plot $t^{0.52}C_{e}(x,t)$ versus $x/t^{0.52}$, and in (f), we plot $t^{0.53}C_{e}(x,t)$ versus $x/t^{0.53}$. We consider $2\times 10^5$ independent realizations for averaging and $N=2048$. This figure demonstrates the unusual scenario where energy correlations can be diffusive ($\lambda=0.5$ and $\lambda = 1.5 $) although the corresponding spin correlations have KPZ behaviour [Fig.~\ref{fig:spin}].}
	\label{fig:ener}
\end{figure}

We consider three cases for the \textit{ip}ILLL model: $\lambda=0.1, 0.5, 1.5$ which approximately falls under perturbative, intermediate and highly non-perturbative parameter regimes respectively. Below, we summarize our results. 

\begin{center}
	\begin{figure}[htbp!]
		\begin{subfigure}{1\linewidth}
			\includegraphics[width=1\linewidth]{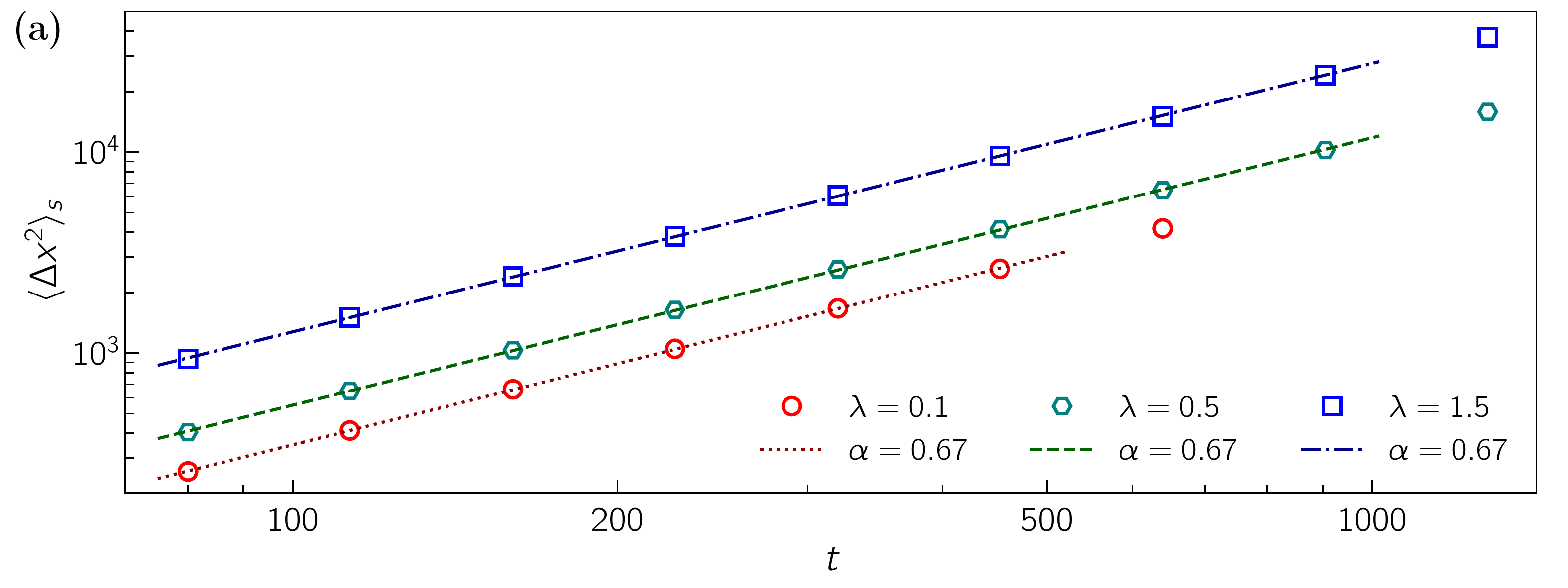}
		\end{subfigure}
		\begin{subfigure}{1\linewidth}
			\includegraphics[width=1\linewidth]{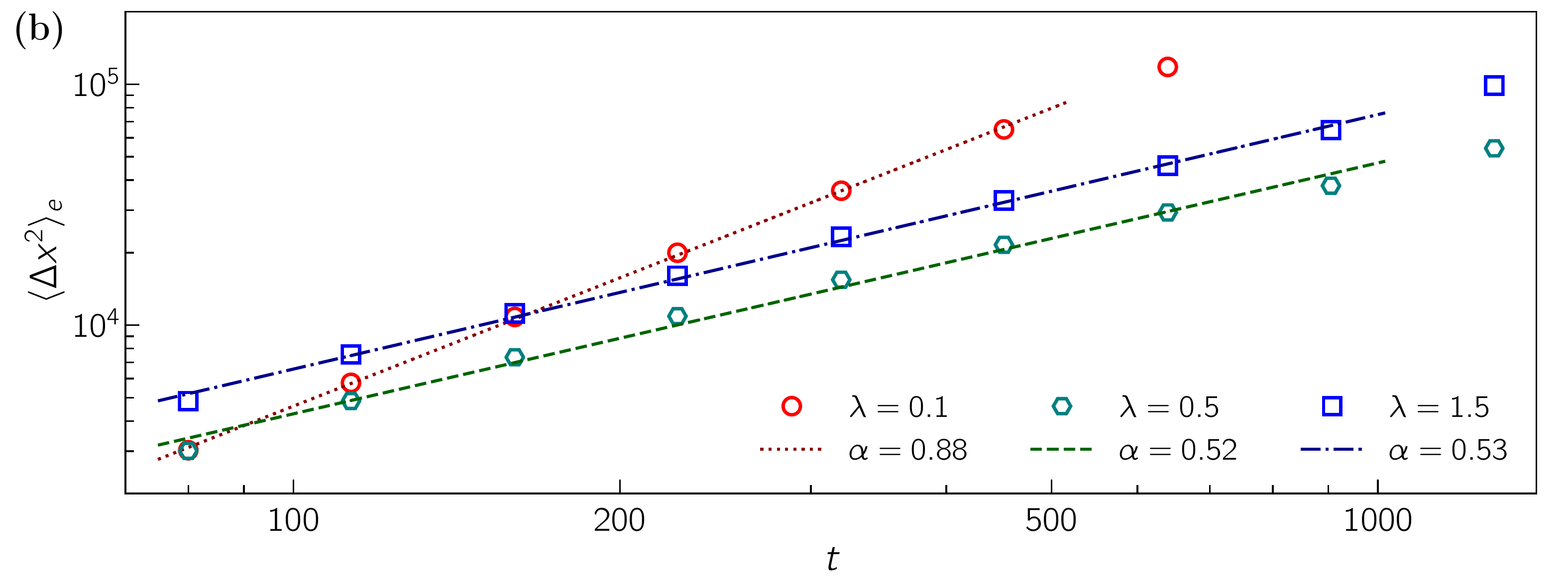}
		\end{subfigure}
		\caption{(Color online) Plots of the MSDs for (a) spin and (b) energy for $\lambda=0.1, 0.5,1.5$. We see remarkable robustness of KPZ behaviour in spin correlation even when integrability-breaking is significant. The last data points are omitted during fitting process to avoid potential boundary effects.}
		\label{fig:msd}
	\end{figure}
\end{center}

\textit{Weakly perturbative regime:} When $\lambda=0.1$ the strength of the integrability breaking term is relatively weak. Nonetheless, we find that although the system is still chaotic, even at significantly long times, the integrable part dominates over the perturbation and the KPZ superdiffusion observed in the integrable case ($\lambda=0$)~\cite{2019-das--dhar} survives in this case as well. This is a surprising result in itself and is consistent with similar predictions in the analogous quantum case~\cite{2021-nardis--ware}. We observe KPZ superdiffusion for spin transport and ballistic transport for energy upto time $t=640$. We show the correlation for spin in Fig.~\ref{fig:spin}(a) and its remarkable collapse when scaled with the KPZ exponent $\alpha = 2/3$ (inset). In Fig.~\ref{fig:spin}(b), we plot the scaled function on a logarithmic scale to show that we see agreement with not only the KPZ exponent but also with the Pr\"{a}hofer-Spohn KPZ scaling function~\cite{2004-prahofer--spohn}. 
The correlation for energy has a scaling exponent $\alpha \approx 0.9$ and exhibits two ballistically moving peaks [see Fig.~\ref{fig:ener}(a) and (b)]. We plot the MSDs for spin and energy correlation in Fig.~\ref{fig:msd}(a) and (b) respectively. Using a linear fit, we obtain $\alpha \approx 0.67$ for spin consistent with the scaling in Fig.~\ref{fig:spin}(b). Similarly, we find $\alpha \approx 0.88$ for energy correlation using a linear fit in Fig.~\ref{fig:msd}(b) which is close to the scaling in Fig.~\ref{fig:ener}(b).

\textit{Intermediate regime:} When $\lambda=0.5$, the system is in the intermediate coupling regime (non-pertubative) where one would expect significant impact of integrability-breaking terms. However, remarkably in this case too, we observe that KPZ superdiffusion prevails for the spin transport. We plot the spin correlation in Fig.~\ref{fig:spin}(c) and (d). The corresponding MSD [Fig.~\ref{fig:msd}~(a)] gives the exponent $\alpha \approx 0.66$ which confirms the scaling in Fig.~\ref{fig:spin}(d). The energy correlation exhibits diffusive behaviour for long times in this case [see Fig.~\ref{fig:ener}(c) and (d)]. This is supported by the computation of the MSD for energy [see Fig.~\ref{fig:msd}(d)]  where we obtain $\alpha \approx 0.52$ for $t > 80$. It is worth noting that this is a very unusual scenario in which a model exhibits diffusive behaviour in energy but KPZ superdiffusion in spin correlations.

\textit{Highly non-perturbative regime:}
To investigate how robust the KPZ behaviour is, we further ramp up the contribution of the integrability breaking term. We consider $\lambda=1.5$, where the energy contribution of the integrability-breaking term is even greater than twice that of the integrable term. To our suprise, we observe KPZ superdiffusion for spin transport in this case too. We show the KPZ scaling of the spin correlation in Fig.~\ref{fig:spin}(f). As in the other cases, we also compare with the exact KPZ scaling function and find good agreement. The energy transport is diffusive in this case as well [see Figs.~\ref{fig:ener}(e) and (f)]. The MSDs (see Fig.~\ref{fig:msd}) for the spin and energy correlation give the values $\alpha \approx 0.67$ and $\alpha \approx 0.53$ respectively consistent with the scalings, in Fig.~\ref{fig:spin}(f) and Fig.~\ref{fig:ener}(f).

Thus, our results show that as long as the integrability-breaking term in the Hamiltonian is isotropic, the spin transport shows KPZ scaling. Although our results correspond to $\beta=1$, our observation should hold true at any temperature. When we consider integrability breaking terms that do not respect spin symmetry, then we immediately find deviations from KPZ behaviour~\cite{2022-supp}. We also consider different types of integrability-breaking but spin-symmetry preserving terms and our computations indicate that the robustness of KPZ behaviour holds at least for a wide family of models~\cite{2022-supp}.

\begin{center}
	\begin{figure}[htbp!]
			\begin{subfigure}{0.5\linewidth}
				\includegraphics[width=1\linewidth]{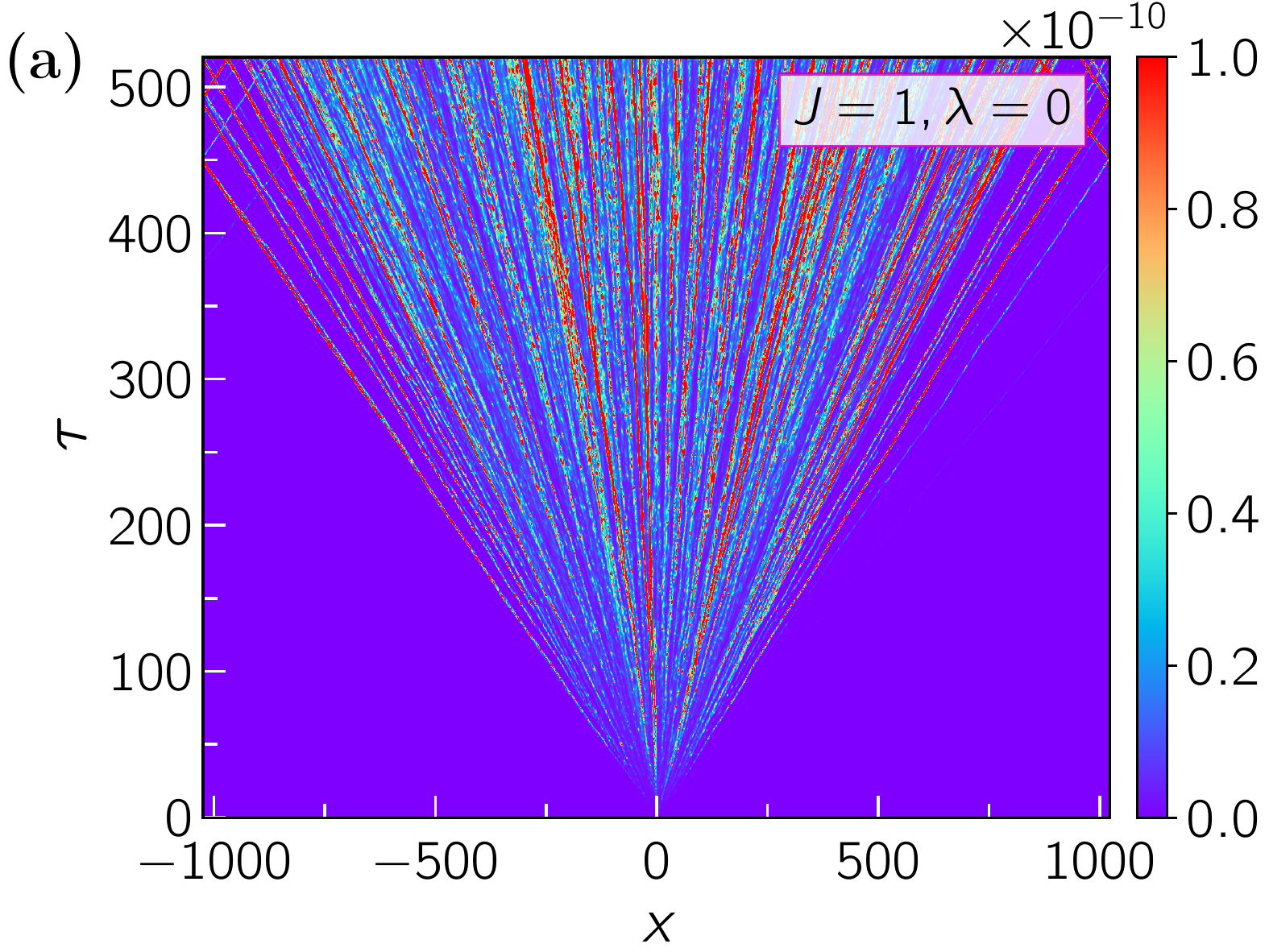}
			\end{subfigure}%
			\begin{subfigure}{0.5\linewidth}
				\includegraphics[width=1\linewidth]{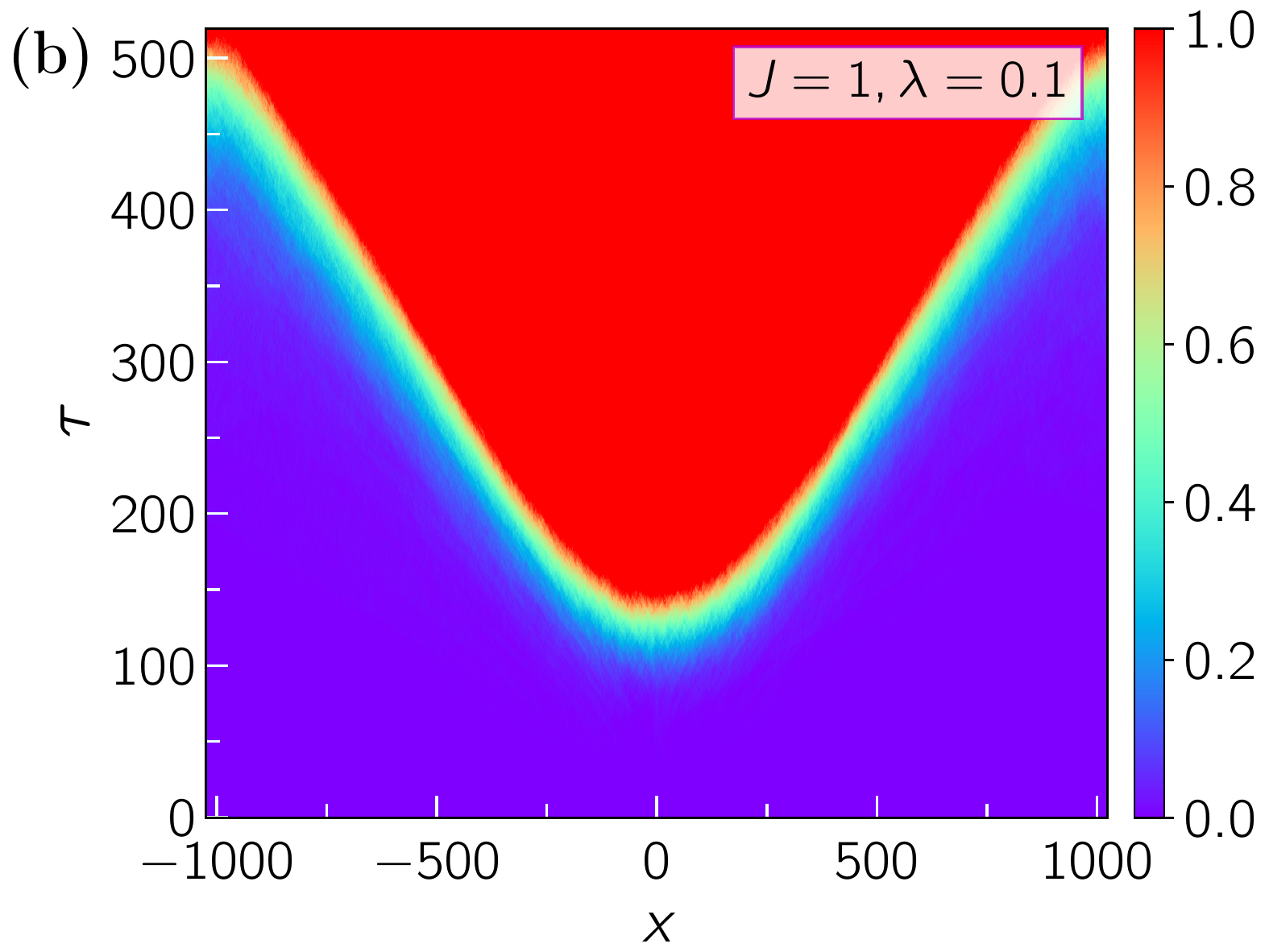}
			\end{subfigure}
			\begin{subfigure}{0.5\linewidth}
				\includegraphics[width=1\linewidth]{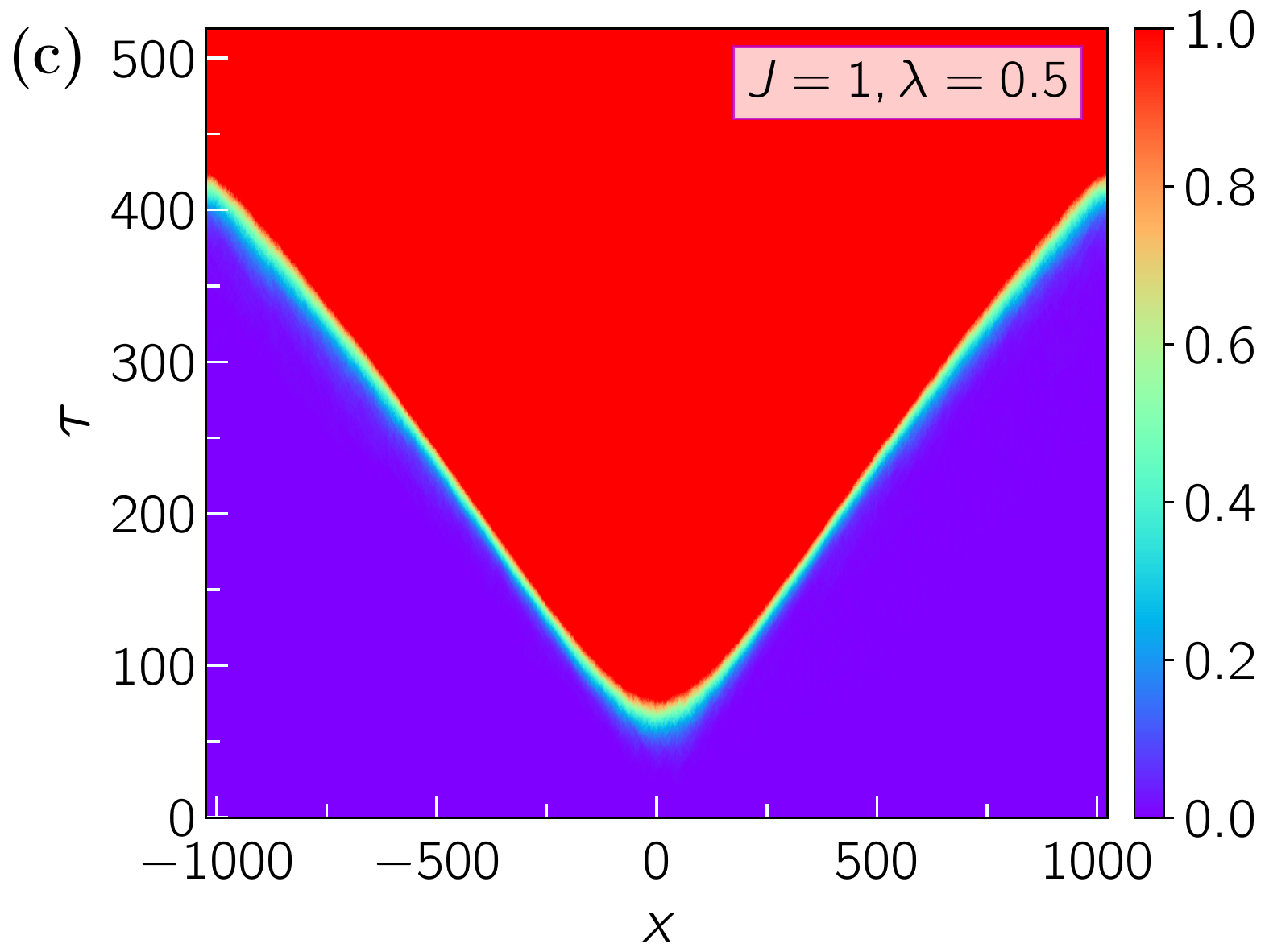}
			\end{subfigure}%
			\begin{subfigure}{0.5\linewidth}
				\includegraphics[width=1\linewidth]{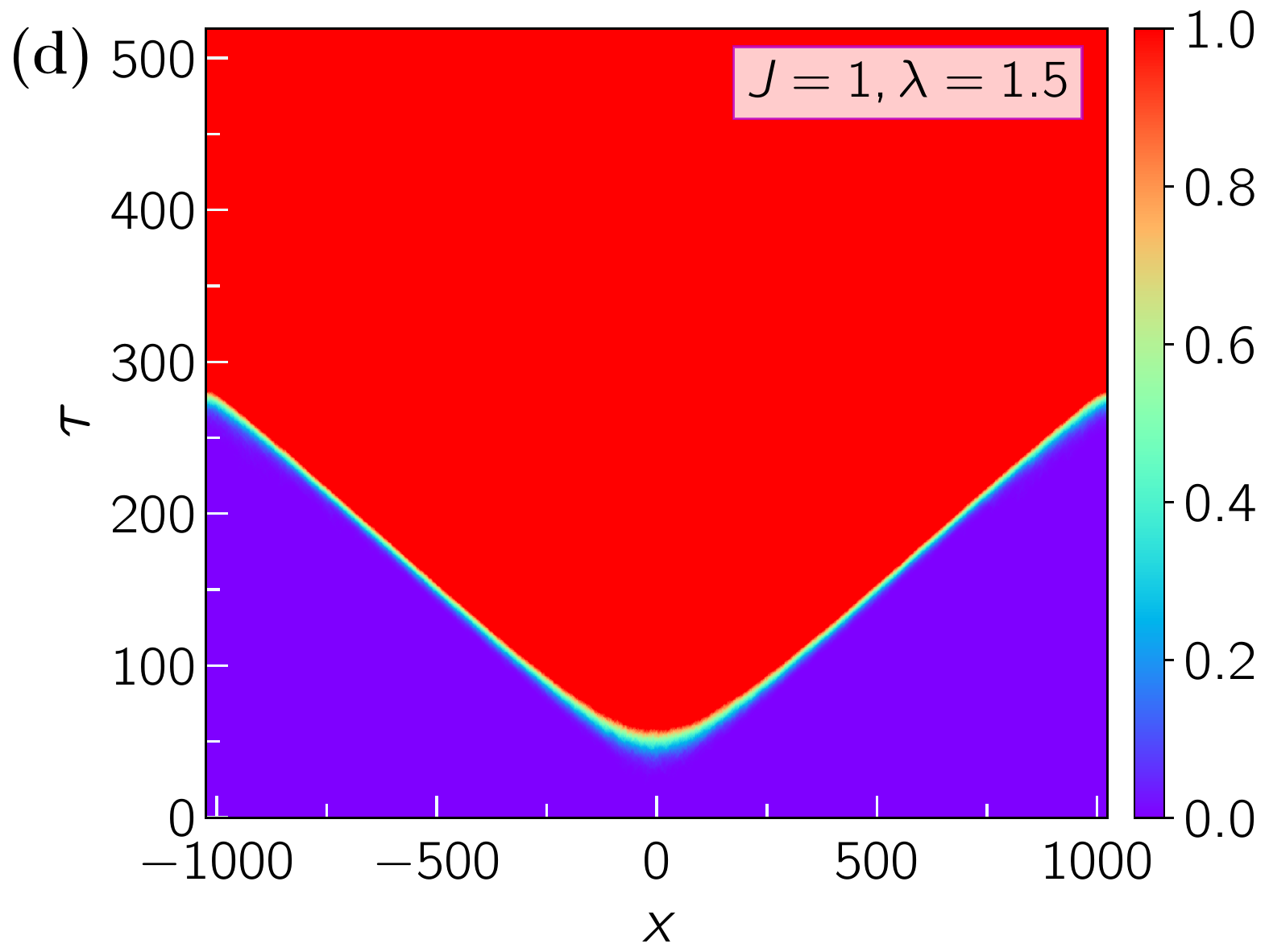}
			\end{subfigure}
			\begin{subfigure}{1\linewidth}
				\includegraphics[width=1\linewidth]{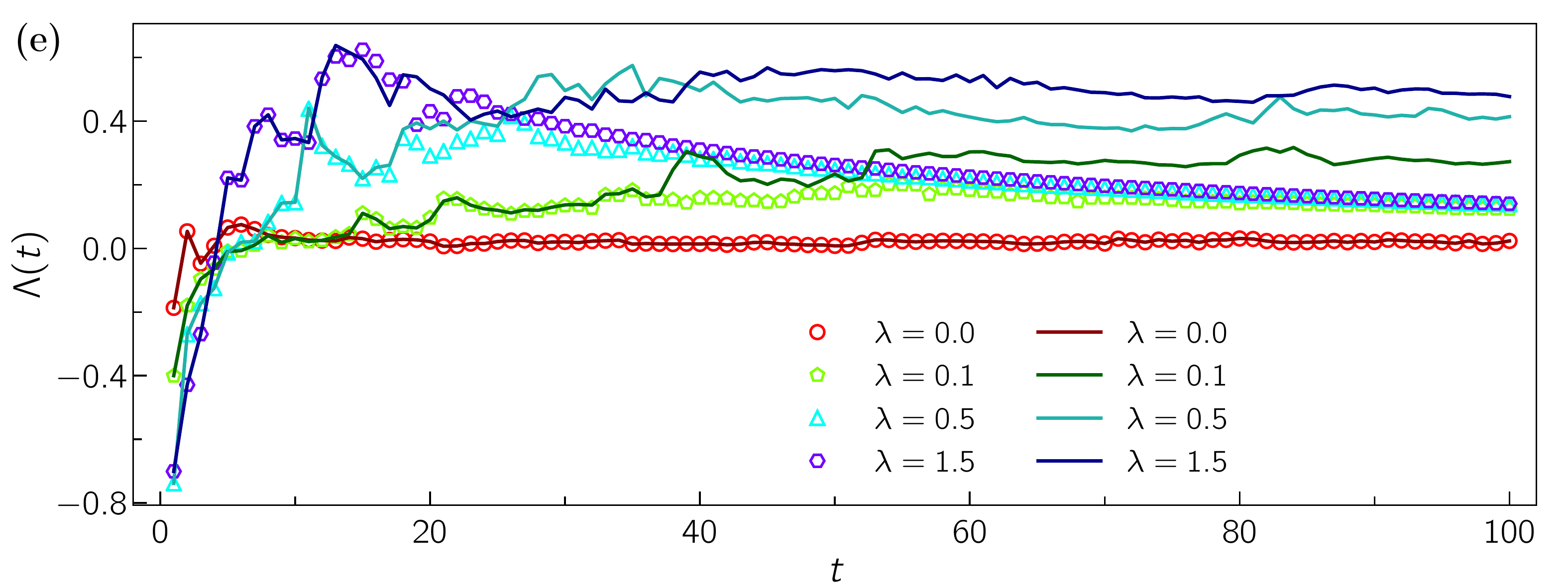}
			\end{subfigure}
			\caption{(Color online) Heatmaps of the OTOC for different values of $\lambda$ with $\epsilon =10^{-6}$. We plot $D(x,t)$ for $\lambda=0$ in (a), $\lambda=0.1$ in (b), $\lambda=0.5$ in (c), and $\lambda=1.5$ in (d). In (e), we show the time variation of the Lyapunov exponent $\Lambda_D(t)$ and $\Lambda_L(t)$ (solid lines). We averaged over $1000$ independent realizations to compute $D(x,t)$ with $N=2048$ and $\Lambda(t)$ with $N=512$.
			}
			\label{fig:hmap}
	\end{figure}
\end{center}

One might wonder if the robustness in KPZ behaviour even when integrability is broken ($\lambda\neq 0 $) is rooted in the fact that the final system is still close to integrable (non-chaotic). To rule out this possibility, we demonstrate that the system is chaotic as soon as $\lambda$ coupling is turned on. To do so, we compute the out-of-time-ordered correlator (OTOC) in the \textit{ip}ILLL model. The OTOC has recently been studied in several classical models as a diagnostic tool to probe how initially localized perturbations spread spatially and grow (or decay) temporally~\cite{2018-das--bhattacharjee, 2018-khemani--nahum, 2018-bilitweski--moessner, 2018-jalabert--wisniacki, 2019-chavez--hirsch, 2020-chatterjee--kulkarni, 2020-kumar--dhar, 2021-ruidas-banerjee, 2021-bilitewski--moessner, 2021-s--kulkarni}. In order to compute the OTOC for the spin chains, we consider the following scheme. From an equilibrium initial configuration, which we denote by A, we generate a \emph{perturbed} copy B by replacing the $N/2$-th spin  with $
	\vec{S}_{N/2}' = ( \vec{S}_{N/2} + \vec{p}_{\epsilon} )/{ | \vec{S}_{N/2} + \vec{p}_{\epsilon}  | }
$
where $\vec{p}_{\epsilon}  = (0,0, \epsilon), \epsilon > 0$. We evolve the 
two copies A and B and compute the OTOC defined as~\cite{2018-das--bhattacharjee} 
\begin{equation}
	D(x,t) = 2 \left( 1 - \langle \vec{S}_{N/2+x}^{\, A}(t)  \cdot  \vec{S}_{N/2 + x}^{ \, B}(t) \rangle \right)  ,
\end{equation}
%\textcolor{red}{stop}.
\\
where $\vec{S}_{n}^{\,A}(t)$ [resp. $\vec{S}_{n}^{\,B}(t)$] is the spin at site $n$ in the copy A (resp. copy B) of the spin chain. In connection with the OTOC we define the finite-time Lyapunov exponent as $\Lambda_D(t) = \ln| D(0,t) / \epsilon^2|/ 2 t$. We also find a linearized equation for $\delta \vec{S}_{n} = \vec{S}_{n}^{\,B}(t) - \vec{S}_{n}^{\,A}(t)$ with $\epsilon \rightarrow 0$~\cite{2022-supp}. In terms of $\delta \vec{S}_{n}$, the Lyapunov exponent is $\Lambda_L(t) = \ln| \langle \delta \vec{S}_{N/2}^2 \rangle / \epsilon^2|/ 2 t$. We show the OTOC for the integrable case ($\lambda=0$) as well as the three cases mentioned above in Figs.~\ref{fig:hmap}(a)-(d) in the form of heatmaps. These heatmaps show nontrivial behaviour as soon as $\lambda$ is turned on. The OTOC in Figs.~\ref{fig:hmap}(b)-(d) indicate the presence of chaotic behaviour, as expected when integrability is broken. The behaviour of the OTOC in \textit{ip}ILLL model resembles that for the classical Heisenberg model~\cite{2018-das--bhattacharjee}.
Even for small $\lambda$, the system becomes significantly chaotic. We note that the butterfly velocity (slope of the cone) increases as we increase the strength of the integrability-breaking term.  In Fig.~\ref{fig:hmap} (e) we show the finite-time Lyapunov exponent $\Lambda(t)$ [both $\Lambda_D(t)$ and $\Lambda_L(t)$] as a function of time. It is clear that $\Lambda(t \rightarrow \infty)\approx 0$ when $\lambda=0$. As soon as we turn on the integrability breaking term ($\lambda\neq 0$), we see that $\Lambda(t \rightarrow \infty)$ is positive and increases with $\lambda$ thereby indicating chaos. Note that $\Lambda_D(t)$ and $\Lambda_L(t)$ agree at early times, while at late times $\Lambda_D(t)$ shows a decay ($\sim 1/t$), as expected for any finite $\epsilon$.

In conclusion, we have studied transport properties in the presence of integrablity-breaking perturbation in a classical spin chain, namely the \textit{ip}LLL model. Our numerical investigation establishes the robustness of the KPZ physics for spin correlations under spin-symmetry preserving but integrability-breaking perturbations of the integrable Hamiltonian. The robustness of KPZ behaviour remains even in the highly non-perturbative regime. In the limit $\lambda / J  \gg 1 $, however, we expect that the features of the classical Heisenberg spin chain will take over for the spin transport at long times destroying the KPZ superdiffusion. In this limit we expect diffusive behaviour with possible logarithmic corrections~\cite{1990-gerling-landau, 1991-liu--muller, 1992-bonfim-reiter, 1993-bohm--leschke, 1994-lovesey-balcar, 1994-srivastava--muller, 2001-meister--sanchez, 2013-bagchi, 2018-das--bhattacharjee, 2019-li, 2020-nardis--ilievski, 2021-glorioso--lucas, 2022-mcroberts--moessner}. For integrability-breaking perturbations which do not respect spin-symmetry, the KPZ superdiffusion is immediately lost~\cite{2022-supp}. Our findings on classical spin chains strongly support the corresponding results for quantum systems~\cite{2021-nardis--ware} and predict the possible robustness of KPZ physics in quantum models even deep in the non-perturbative regime. Despite this robustness to integrability-breaking terms, one cannot rule out the crossover to features finally dominated by non-integrable terms (such as conventional diffusion~\cite{ 2020-dupont-moore, 2021-glorioso--lucas, 2021-claeys--arbeitman}) at extremely long times and large system sizes inaccessible in present state-of-the-art computations.

The KPZ scaling in non-integrable but spin-symmetry preserving systems could be rooted in a  possible robustness of solitons of the ILLL in the presence of integrability-breaking but spin symmetry preserving terms and this will be explored in future. The anisotropic but integrable generalization of ILLL and the effect of breaking its integrability is an interesting question that is expected to yield a plethora of possibilities.  \\

\textit{Acknowledgements:}
We thank A. Das for useful discussions. MK would like to acknowledge support from the project 6004-1 of the Indo-French Centre for the Pro- motion of Advanced Research (IFCPAR), Ramanujan Fellowship (SB/S2/RJN-114/2016), SERB Early Career Research Award (ECR/2018/002085) and SERB Matrics Grant (MTR/2019/001101) from the Science and En- gineering Research Board (SERB), Department of Science and Technology, Government of India. DR, AD and MK acknowledge support of the Department of Atomic Energy, Government of India, under Project No. RTI4001.
 This research was supported in part by the International Centre for Theoretical Sciences (ICTS) for the online program - Hydrodynamics and fluctuations - microscopic approaches in condensed matter systems (code: ICTS/hydro2021/9). 

\appendix

\bibliographystyle{apsrev4-1}
\bibliography{refs.bib}

%merlin.mbs apsrev4-1.bst 2010-07-25 4.21a (PWD, AO, DPC) hacked
%Control: key (0)
%Control: author (72) initials jnrlst
%Control: editor formatted (1) identically to author
%Control: production of article title (-1) disabled
%Control: page (0) single
%Control: year (1) truncated
%Control: production of eprint (0) enabled
\begin{thebibliography}{48}%
\makeatletter
\providecommand \@ifxundefined [1]{%
 \@ifx{#1\undefined}
}%
\providecommand \@ifnum [1]{%
 \ifnum #1\expandafter \@firstoftwo
 \else \expandafter \@secondoftwo
 \fi
}%
\providecommand \@ifx [1]{%
 \ifx #1\expandafter \@firstoftwo
 \else \expandafter \@secondoftwo
 \fi
}%
\providecommand \natexlab [1]{#1}%
\providecommand \enquote  [1]{``#1''}%
\providecommand \bibnamefont  [1]{#1}%
\providecommand \bibfnamefont [1]{#1}%
\providecommand \citenamefont [1]{#1}%
\providecommand \href@noop [0]{\@secondoftwo}%
\providecommand \href [0]{\begingroup \@sanitize@url \@href}%
\providecommand \@href[1]{\@@startlink{#1}\@@href}%
\providecommand \@@href[1]{\endgroup#1\@@endlink}%
\providecommand \@sanitize@url [0]{\catcode `\\12\catcode `\$12\catcode
  `\&12\catcode `\#12\catcode `\^12\catcode `\_12\catcode `\%12\relax}%
\providecommand \@@startlink[1]{}%
\providecommand \@@endlink[0]{}%
\providecommand \url  [0]{\begingroup\@sanitize@url \@url }%
\providecommand \@url [1]{\endgroup\@href {#1}{\urlprefix }}%
\providecommand \urlprefix  [0]{URL }%
\providecommand \Eprint [0]{\href }%
\providecommand \doibase [0]{http://dx.doi.org/}%
\providecommand \selectlanguage [0]{\@gobble}%
\providecommand \bibinfo  [0]{\@secondoftwo}%
\providecommand \bibfield  [0]{\@secondoftwo}%
\providecommand \translation [1]{[#1]}%
\providecommand \BibitemOpen [0]{}%
\providecommand \bibitemStop [0]{}%
\providecommand \bibitemNoStop [0]{.\EOS\space}%
\providecommand \EOS [0]{\spacefactor3000\relax}%
\providecommand \BibitemShut  [1]{\csname bibitem#1\endcsname}%
\let\auto@bib@innerbib\@empty
%</preamble>
\bibitem [{\citenamefont {\ifmmode \check{Z}\else
  \v{Z}\fi{}nidari\ifmmode~\check{c}\else \v{c}\fi{}}(2011)}]{2011-znidaric}%
  \BibitemOpen
  \bibfield  {author} {\bibinfo {author} {\bibfnamefont {M.}~\bibnamefont
  {\ifmmode \check{Z}\else \v{Z}\fi{}nidari\ifmmode~\check{c}\else
  \v{c}\fi{}}},\ }\href {\doibase 10.1103/PhysRevLett.106.220601} {\bibfield
  {journal} {\bibinfo  {journal} {Phys. Rev. Lett.}\ }\textbf {\bibinfo
  {volume} {106}},\ \bibinfo {pages} {220601} (\bibinfo {year}
  {2011})}\BibitemShut {NoStop}%
\bibitem [{\citenamefont {Ljubotina}\ \emph {et~al.}(2017)\citenamefont
  {Ljubotina}, \citenamefont {\ifmmode
  \check{Z}\else\v{Z}\fi{}nidari\ifmmode~\check{c}\else \v{c}\fi{}},\ and\
  \citenamefont {Prosen}}]{2017-ljubotina--prosen}%
  \BibitemOpen
  \bibfield  {author} {\bibinfo {author} {\bibfnamefont {M.}~\bibnamefont
  {Ljubotina}}, \bibinfo {author} {\bibfnamefont {M.}~\bibnamefont {\ifmmode
  \check{Z}\else\v{Z}\fi{}nidari\ifmmode~\check{c}\else \v{c}\fi{}}}, \ and\
  \bibinfo {author} {\bibfnamefont {T.}~\bibnamefont {Prosen}},\ }\href
  {\doibase 10.1038/ncomms16117} {\bibfield  {journal} {\bibinfo  {journal}
  {Nature Communications}\ }\textbf {\bibinfo {volume} {8}},\ \bibinfo {pages}
  {16117} (\bibinfo {year} {2017})}\BibitemShut {NoStop}%
\bibitem [{\citenamefont {Ljubotina}\ \emph {et~al.}(2019)\citenamefont
  {Ljubotina}, \citenamefont
  {\ifmmode\check{Z}\else\v{Z}\fi{}nidari\ifmmode\check{c}\else\v{c}\fi{}},\
  and\ \citenamefont {Prosen}}]{2019-ljubotina--prosen}%
  \BibitemOpen
  \bibfield  {author} {\bibinfo {author} {\bibfnamefont {M.}~\bibnamefont
  {Ljubotina}}, \bibinfo {author} {\bibfnamefont {M.}~\bibnamefont
  {\ifmmode\check{Z}\else\v{Z}\fi{}nidari\ifmmode\check{c}\else\v{c}\fi{}}}, \
  and\ \bibinfo {author} {\bibfnamefont {T.}~\bibnamefont {Prosen}},\ }\href
  {\doibase 10.1103/PhysRevLett.122.210602} {\bibfield  {journal} {\bibinfo
  {journal} {Phys. Rev. Lett.}\ }\textbf {\bibinfo {volume} {122}},\ \bibinfo
  {pages} {210602} (\bibinfo {year} {2019})}\BibitemShut {NoStop}%
\bibitem [{\citenamefont {Pr{\"a}hofer}\ and\ \citenamefont
  {Spohn}(2004)}]{2004-prahofer--spohn}%
  \BibitemOpen
  \bibfield  {author} {\bibinfo {author} {\bibfnamefont {M.}~\bibnamefont
  {Pr{\"a}hofer}}\ and\ \bibinfo {author} {\bibfnamefont {H.}~\bibnamefont
  {Spohn}},\ }\href {\doibase 10.1023/B:JOSS.0000019810.21828.fc} {\bibfield
  {journal} {\bibinfo  {journal} {Journal of Statistical Physics}\ }\textbf
  {\bibinfo {volume} {115}},\ \bibinfo {pages} {255} (\bibinfo {year}
  {2004})}\BibitemShut {NoStop}%
\bibitem [{\citenamefont {Kardar}\ \emph {et~al.}(1986)\citenamefont {Kardar},
  \citenamefont {Parisi},\ and\ \citenamefont {Zhang}}]{1986-kardar--zhang}%
  \BibitemOpen
  \bibfield  {author} {\bibinfo {author} {\bibfnamefont {M.}~\bibnamefont
  {Kardar}}, \bibinfo {author} {\bibfnamefont {G.}~\bibnamefont {Parisi}}, \
  and\ \bibinfo {author} {\bibfnamefont {Y.-C.}\ \bibnamefont {Zhang}},\ }\href
  {\doibase 10.1103/PhysRevLett.56.889} {\bibfield  {journal} {\bibinfo
  {journal} {Phys. Rev. Lett.}\ }\textbf {\bibinfo {volume} {56}},\ \bibinfo
  {pages} {889} (\bibinfo {year} {1986})}\BibitemShut {NoStop}%
\bibitem [{\citenamefont {Takeuchi}(2018)}]{2018-takeuchi}%
  \BibitemOpen
  \bibfield  {author} {\bibinfo {author} {\bibfnamefont {K.~A.}\ \bibnamefont
  {Takeuchi}},\ }\href {\doibase https://doi.org/10.1016/j.physa.2018.03.009}
  {\bibfield  {journal} {\bibinfo  {journal} {Physica A: Statistical Mechanics
  and its Applications}\ }\textbf {\bibinfo {volume} {504}},\ \bibinfo {pages}
  {77 } (\bibinfo {year} {2018})},\ \bibinfo {note} {{Lecture Notes of the 14th
  International Summer School on Fundamental Problems in Statistical
  Physics}}\BibitemShut {NoStop}%
\bibitem [{\citenamefont {Dupont}\ and\ \citenamefont
  {Moore}(2020)}]{2020-dupont-moore}%
  \BibitemOpen
  \bibfield  {author} {\bibinfo {author} {\bibfnamefont {M.}~\bibnamefont
  {Dupont}}\ and\ \bibinfo {author} {\bibfnamefont {J.~E.}\ \bibnamefont
  {Moore}},\ }\href {\doibase 10.1103/PhysRevB.101.121106} {\bibfield
  {journal} {\bibinfo  {journal} {Phys. Rev. B}\ }\textbf {\bibinfo {volume}
  {101}},\ \bibinfo {pages} {121106} (\bibinfo {year} {2020})}\BibitemShut
  {NoStop}%
\bibitem [{\citenamefont {Castro-Alvaredo}\ \emph {et~al.}(2016)\citenamefont
  {Castro-Alvaredo}, \citenamefont {Doyon},\ and\ \citenamefont
  {Yoshimura}}]{2016-alvaredo--yoshimura}%
  \BibitemOpen
  \bibfield  {author} {\bibinfo {author} {\bibfnamefont {O.~A.}\ \bibnamefont
  {Castro-Alvaredo}}, \bibinfo {author} {\bibfnamefont {B.}~\bibnamefont
  {Doyon}}, \ and\ \bibinfo {author} {\bibfnamefont {T.}~\bibnamefont
  {Yoshimura}},\ }\href {\doibase 10.1103/PhysRevX.6.041065} {\bibfield
  {journal} {\bibinfo  {journal} {Phys. Rev. X}\ }\textbf {\bibinfo {volume}
  {6}},\ \bibinfo {pages} {041065} (\bibinfo {year} {2016})}\BibitemShut
  {NoStop}%
\bibitem [{\citenamefont {Ilievski}\ \emph {et~al.}(2018)\citenamefont
  {Ilievski}, \citenamefont {De~Nardis}, \citenamefont {Medenjak},\ and\
  \citenamefont {Prosen}}]{2018-ilievski--prosen}%
  \BibitemOpen
  \bibfield  {author} {\bibinfo {author} {\bibfnamefont {E.}~\bibnamefont
  {Ilievski}}, \bibinfo {author} {\bibfnamefont {J.}~\bibnamefont {De~Nardis}},
  \bibinfo {author} {\bibfnamefont {M.}~\bibnamefont {Medenjak}}, \ and\
  \bibinfo {author} {\bibfnamefont {T.}~\bibnamefont {Prosen}},\ }\href
  {\doibase 10.1103/PhysRevLett.121.230602} {\bibfield  {journal} {\bibinfo
  {journal} {Phys. Rev. Lett.}\ }\textbf {\bibinfo {volume} {121}},\ \bibinfo
  {pages} {230602} (\bibinfo {year} {2018})}\BibitemShut {NoStop}%
\bibitem [{\citenamefont {Gopalakrishnan}\ and\ \citenamefont
  {Vasseur}(2019)}]{2019-gopalakrishnan-vasseur}%
  \BibitemOpen
  \bibfield  {author} {\bibinfo {author} {\bibfnamefont {S.}~\bibnamefont
  {Gopalakrishnan}}\ and\ \bibinfo {author} {\bibfnamefont {R.}~\bibnamefont
  {Vasseur}},\ }\href {\doibase 10.1103/PhysRevLett.122.127202} {\bibfield
  {journal} {\bibinfo  {journal} {Phys. Rev. Lett.}\ }\textbf {\bibinfo
  {volume} {122}},\ \bibinfo {pages} {127202} (\bibinfo {year}
  {2019})}\BibitemShut {NoStop}%
\bibitem [{\citenamefont {De~Nardis}\ \emph
  {et~al.}(2020{\natexlab{a}})\citenamefont {De~Nardis}, \citenamefont
  {Gopalakrishnan}, \citenamefont {Ilievski},\ and\ \citenamefont
  {Vasseur}}]{2020-nardis--vasseur}%
  \BibitemOpen
  \bibfield  {author} {\bibinfo {author} {\bibfnamefont {J.}~\bibnamefont
  {De~Nardis}}, \bibinfo {author} {\bibfnamefont {S.}~\bibnamefont
  {Gopalakrishnan}}, \bibinfo {author} {\bibfnamefont {E.}~\bibnamefont
  {Ilievski}}, \ and\ \bibinfo {author} {\bibfnamefont {R.}~\bibnamefont
  {Vasseur}},\ }\href {\doibase 10.1103/PhysRevLett.125.070601} {\bibfield
  {journal} {\bibinfo  {journal} {Phys. Rev. Lett.}\ }\textbf {\bibinfo
  {volume} {125}},\ \bibinfo {pages} {070601} (\bibinfo {year}
  {2020}{\natexlab{a}})}\BibitemShut {NoStop}%
\bibitem [{\citenamefont {Ilievski}\ \emph {et~al.}(2021)\citenamefont
  {Ilievski}, \citenamefont {De~Nardis}, \citenamefont {Gopalakrishnan},
  \citenamefont {Vasseur},\ and\ \citenamefont {Ware}}]{2021-ilievski--ware}%
  \BibitemOpen
  \bibfield  {author} {\bibinfo {author} {\bibfnamefont {E.}~\bibnamefont
  {Ilievski}}, \bibinfo {author} {\bibfnamefont {J.}~\bibnamefont {De~Nardis}},
  \bibinfo {author} {\bibfnamefont {S.}~\bibnamefont {Gopalakrishnan}},
  \bibinfo {author} {\bibfnamefont {R.}~\bibnamefont {Vasseur}}, \ and\
  \bibinfo {author} {\bibfnamefont {B.}~\bibnamefont {Ware}},\ }\href {\doibase
  10.1103/PhysRevX.11.031023} {\bibfield  {journal} {\bibinfo  {journal} {Phys.
  Rev. X}\ }\textbf {\bibinfo {volume} {11}},\ \bibinfo {pages} {031023}
  (\bibinfo {year} {2021})}\BibitemShut {NoStop}%
\bibitem [{\citenamefont {Bulchandani}\ \emph {et~al.}(2021)\citenamefont
  {Bulchandani}, \citenamefont {Gopalakrishnan},\ and\ \citenamefont
  {Ilievski}}]{2021-bulchandani--ilievski}%
  \BibitemOpen
  \bibfield  {author} {\bibinfo {author} {\bibfnamefont {V.~B.}\ \bibnamefont
  {Bulchandani}}, \bibinfo {author} {\bibfnamefont {S.}~\bibnamefont
  {Gopalakrishnan}}, \ and\ \bibinfo {author} {\bibfnamefont {E.}~\bibnamefont
  {Ilievski}},\ }\href {\doibase 10.1088/1742-5468/ac12c7} {\bibfield
  {journal} {\bibinfo  {journal} {Journal of Statistical Mechanics: Theory and
  Experiment}\ }\textbf {\bibinfo {volume} {2021}},\ \bibinfo {pages} {084001}
  (\bibinfo {year} {2021})}\BibitemShut {NoStop}%
\bibitem [{\citenamefont {Scheie}\ \emph {et~al.}(2021)\citenamefont {Scheie},
  \citenamefont {Sherman}, \citenamefont {Dupont}, \citenamefont {Nagler},
  \citenamefont {Stone}, \citenamefont {Granroth}, \citenamefont {Moore},\ and\
  \citenamefont {Tennant}}]{2021-scheie--tennant}%
  \BibitemOpen
  \bibfield  {author} {\bibinfo {author} {\bibfnamefont {A.}~\bibnamefont
  {Scheie}}, \bibinfo {author} {\bibfnamefont {N.~E.}\ \bibnamefont {Sherman}},
  \bibinfo {author} {\bibfnamefont {M.}~\bibnamefont {Dupont}}, \bibinfo
  {author} {\bibfnamefont {S.~E.}\ \bibnamefont {Nagler}}, \bibinfo {author}
  {\bibfnamefont {M.~B.}\ \bibnamefont {Stone}}, \bibinfo {author}
  {\bibfnamefont {G.~E.}\ \bibnamefont {Granroth}}, \bibinfo {author}
  {\bibfnamefont {J.~E.}\ \bibnamefont {Moore}}, \ and\ \bibinfo {author}
  {\bibfnamefont {D.~A.}\ \bibnamefont {Tennant}},\ }\href {\doibase
  10.1038/s41567-021-01191-6} {\bibfield  {journal} {\bibinfo  {journal}
  {Nature Physics}\ }\textbf {\bibinfo {volume} {17}},\ \bibinfo {pages} {726}
  (\bibinfo {year} {2021})}\BibitemShut {NoStop}%
\bibitem [{\citenamefont {Wei}\ \emph {et~al.}(2021)\citenamefont {Wei},
  \citenamefont {Rubio-Abadal}, \citenamefont {Ye}, \citenamefont {Machado},
  \citenamefont {Kemp}, \citenamefont {Srakaew}, \citenamefont {Hollerith},
  \citenamefont {Rui}, \citenamefont {Gopalakrishnan}, \citenamefont {Yao},
  \citenamefont {Bloch},\ and\ \citenamefont {Zeiher}}]{2021-wei--zeiher}%
  \BibitemOpen
  \bibfield  {author} {\bibinfo {author} {\bibfnamefont {D.}~\bibnamefont
  {Wei}}, \bibinfo {author} {\bibfnamefont {A.}~\bibnamefont {Rubio-Abadal}},
  \bibinfo {author} {\bibfnamefont {B.}~\bibnamefont {Ye}}, \bibinfo {author}
  {\bibfnamefont {F.}~\bibnamefont {Machado}}, \bibinfo {author} {\bibfnamefont
  {J.}~\bibnamefont {Kemp}}, \bibinfo {author} {\bibfnamefont {K.}~\bibnamefont
  {Srakaew}}, \bibinfo {author} {\bibfnamefont {S.}~\bibnamefont {Hollerith}},
  \bibinfo {author} {\bibfnamefont {J.}~\bibnamefont {Rui}}, \bibinfo {author}
  {\bibfnamefont {S.}~\bibnamefont {Gopalakrishnan}}, \bibinfo {author}
  {\bibfnamefont {N.~Y.}\ \bibnamefont {Yao}}, \bibinfo {author} {\bibfnamefont
  {I.}~\bibnamefont {Bloch}}, \ and\ \bibinfo {author} {\bibfnamefont
  {J.}~\bibnamefont {Zeiher}},\ }\href {\doibase 10.48550/ARXIV.2107.00038}
  {\enquote {\bibinfo {title} {Quantum gas microscopy of kardar-parisi-zhang
  superdiffusion},}\ } (\bibinfo {year} {2021})\BibitemShut {NoStop}%
\bibitem [{\citenamefont {Prosen}\ and\ \citenamefont {\ifmmode \check{Z}\else
  \v{Z}\fi{}unkovi\ifmmode~\check{c}\else
  \v{c}\fi{}}(2013)}]{2013-prosen-zunkovic}%
  \BibitemOpen
  \bibfield  {author} {\bibinfo {author} {\bibfnamefont {T.}~\bibnamefont
  {Prosen}}\ and\ \bibinfo {author} {\bibfnamefont {B.}~\bibnamefont {\ifmmode
  \check{Z}\else \v{Z}\fi{}unkovi\ifmmode~\check{c}\else \v{c}\fi{}}},\ }\href
  {\doibase 10.1103/PhysRevLett.111.040602} {\bibfield  {journal} {\bibinfo
  {journal} {Phys. Rev. Lett.}\ }\textbf {\bibinfo {volume} {111}},\ \bibinfo
  {pages} {040602} (\bibinfo {year} {2013})}\BibitemShut {NoStop}%
\bibitem [{\citenamefont {Das}\ \emph {et~al.}(2019)\citenamefont {Das},
  \citenamefont {Kulkarni}, \citenamefont {Spohn},\ and\ \citenamefont
  {Dhar}}]{2019-das--dhar}%
  \BibitemOpen
  \bibfield  {author} {\bibinfo {author} {\bibfnamefont {A.}~\bibnamefont
  {Das}}, \bibinfo {author} {\bibfnamefont {M.}~\bibnamefont {Kulkarni}},
  \bibinfo {author} {\bibfnamefont {H.}~\bibnamefont {Spohn}}, \ and\ \bibinfo
  {author} {\bibfnamefont {A.}~\bibnamefont {Dhar}},\ }\href {\doibase
  10.1103/PhysRevE.100.042116} {\bibfield  {journal} {\bibinfo  {journal}
  {Phys. Rev. E}\ }\textbf {\bibinfo {volume} {100}},\ \bibinfo {pages}
  {042116} (\bibinfo {year} {2019})}\BibitemShut {NoStop}%
\bibitem [{\citenamefont {Krajnik}\ and\ \citenamefont
  {Prosen}(2020)}]{2020-krajnik-prosen}%
  \BibitemOpen
  \bibfield  {author} {\bibinfo {author} {\bibfnamefont {{\v Z}.}~\bibnamefont
  {Krajnik}}\ and\ \bibinfo {author} {\bibfnamefont {T.}~\bibnamefont
  {Prosen}},\ }\href {\doibase 10.1007/s10955-020-02523-1} {\bibfield
  {journal} {\bibinfo  {journal} {Journal of Statistical Physics}\ }\textbf
  {\bibinfo {volume} {179}},\ \bibinfo {pages} {110} (\bibinfo {year}
  {2020})}\BibitemShut {NoStop}%
\bibitem [{\citenamefont {Krajnik}\ \emph {et~al.}(2020)\citenamefont
  {Krajnik}, \citenamefont {Ilievski},\ and\ \citenamefont
  {Prosen}}]{2020-krajnik--prosen}%
  \BibitemOpen
  \bibfield  {author} {\bibinfo {author} {\bibfnamefont {{\v Z}.}~\bibnamefont
  {Krajnik}}, \bibinfo {author} {\bibfnamefont {E.}~\bibnamefont {Ilievski}}, \
  and\ \bibinfo {author} {\bibfnamefont {T.}~\bibnamefont {Prosen}},\ }\href
  {\doibase 10.21468/SciPostPhys.9.3.038} {\bibfield  {journal} {\bibinfo
  {journal} {SciPost Phys.}\ }\textbf {\bibinfo {volume} {9}},\ \bibinfo
  {pages} {38} (\bibinfo {year} {2020})}\BibitemShut {NoStop}%
\bibitem [{\citenamefont {De~Nardis}\ \emph {et~al.}(2021)\citenamefont
  {De~Nardis}, \citenamefont {Gopalakrishnan}, \citenamefont {Vasseur},\ and\
  \citenamefont {Ware}}]{2021-nardis--ware}%
  \BibitemOpen
  \bibfield  {author} {\bibinfo {author} {\bibfnamefont {J.}~\bibnamefont
  {De~Nardis}}, \bibinfo {author} {\bibfnamefont {S.}~\bibnamefont
  {Gopalakrishnan}}, \bibinfo {author} {\bibfnamefont {R.}~\bibnamefont
  {Vasseur}}, \ and\ \bibinfo {author} {\bibfnamefont {B.}~\bibnamefont
  {Ware}},\ }\href {\doibase 10.1103/PhysRevLett.127.057201} {\bibfield
  {journal} {\bibinfo  {journal} {Phys. Rev. Lett.}\ }\textbf {\bibinfo
  {volume} {127}},\ \bibinfo {pages} {057201} (\bibinfo {year}
  {2021})}\BibitemShut {NoStop}%
\bibitem [{\citenamefont {Claeys}\ \emph {et~al.}(2021)\citenamefont {Claeys},
  \citenamefont {Lamacraft},\ and\ \citenamefont
  {Herzog-Arbeitman}}]{2021-claeys--arbeitman}%
  \BibitemOpen
  \bibfield  {author} {\bibinfo {author} {\bibfnamefont {P.~W.}\ \bibnamefont
  {Claeys}}, \bibinfo {author} {\bibfnamefont {A.}~\bibnamefont {Lamacraft}}, \
  and\ \bibinfo {author} {\bibfnamefont {J.}~\bibnamefont {Herzog-Arbeitman}},\
  }\href {\doibase 10.48550/ARXIV.2110.06951} {\bibfield  {journal} {\bibinfo
  {journal} {arXiv}\ } (\bibinfo {year} {2021}),\
  10.48550/ARXIV.2110.06951}\BibitemShut {NoStop}%
\bibitem [{\citenamefont {Sklyanin}(1982)}]{1982-sklyanin}%
  \BibitemOpen
  \bibfield  {author} {\bibinfo {author} {\bibfnamefont {E.~K.}\ \bibnamefont
  {Sklyanin}},\ }\href {\doibase 10.1007/BF01077848} {\bibfield  {journal}
  {\bibinfo  {journal} {Functional Analysis and Its Applications}\ }\textbf
  {\bibinfo {volume} {16}},\ \bibinfo {pages} {263} (\bibinfo {year}
  {1982})}\BibitemShut {NoStop}%
\bibitem [{\citenamefont {Sklyanin}(1988)}]{1988-sklyanin}%
  \BibitemOpen
  \bibfield  {author} {\bibinfo {author} {\bibfnamefont {E.~K.}\ \bibnamefont
  {Sklyanin}},\ }\href {\doibase 10.1007/BF01084941} {\bibfield  {journal}
  {\bibinfo  {journal} {Journal of Soviet Mathematics}\ }\textbf {\bibinfo
  {volume} {40}},\ \bibinfo {pages} {93} (\bibinfo {year} {1988})}\BibitemShut
  {NoStop}%
\bibitem [{\citenamefont {Faddeev}\ and\ \citenamefont
  {Takhtajan}(2007)}]{2007-faddeev-takhtajan}%
  \BibitemOpen
  \bibfield  {author} {\bibinfo {author} {\bibfnamefont {L.}~\bibnamefont
  {Faddeev}}\ and\ \bibinfo {author} {\bibfnamefont {L.}~\bibnamefont
  {Takhtajan}},\ }\href {https://www.springer.com/gp/book/9783540698432} {\emph
  {\bibinfo {title} {{Hamiltonian Methods in the Theory of Solitons}}}}\
  (\bibinfo  {publisher} {Springer Science and Business Media, New York},\
  \bibinfo {year} {2007})\BibitemShut {NoStop}%
\bibitem [{202()}]{2022-supp}%
  \BibitemOpen
  \href@noop {} {\bibinfo  {journal} {Supplemental Material}\ }\BibitemShut
  {NoStop}%
\bibitem [{\citenamefont {Das}\ \emph {et~al.}(2020)\citenamefont {Das},
  \citenamefont {Damle}, \citenamefont {Dhar}, \citenamefont {Huse},
  \citenamefont {Kulkarni}, \citenamefont {Mendl},\ and\ \citenamefont
  {Spohn}}]{2020-das--spohn}%
  \BibitemOpen
\bibfield  {journal} {  }\bibfield  {author} {\bibinfo {author} {\bibfnamefont
  {A.}~\bibnamefont {Das}}, \bibinfo {author} {\bibfnamefont {K.}~\bibnamefont
  {Damle}}, \bibinfo {author} {\bibfnamefont {A.}~\bibnamefont {Dhar}},
  \bibinfo {author} {\bibfnamefont {D.~A.}\ \bibnamefont {Huse}}, \bibinfo
  {author} {\bibfnamefont {M.}~\bibnamefont {Kulkarni}}, \bibinfo {author}
  {\bibfnamefont {C.~B.}\ \bibnamefont {Mendl}}, \ and\ \bibinfo {author}
  {\bibfnamefont {H.}~\bibnamefont {Spohn}},\ }\href {\doibase
  10.1007/s10955-019-02397-y} {\bibfield  {journal} {\bibinfo  {journal}
  {Journal of Statistical Physics}\ }\textbf {\bibinfo {volume} {180}},\
  \bibinfo {pages} {238} (\bibinfo {year} {2020})}\BibitemShut {NoStop}%
\bibitem [{\citenamefont {Das}\ \emph {et~al.}(2018)\citenamefont {Das},
  \citenamefont {Chakrabarty}, \citenamefont {Dhar}, \citenamefont {Kundu},
  \citenamefont {Huse}, \citenamefont {Moessner}, \citenamefont {Ray},\ and\
  \citenamefont {Bhattacharjee}}]{2018-das--bhattacharjee}%
  \BibitemOpen
  \bibfield  {author} {\bibinfo {author} {\bibfnamefont {A.}~\bibnamefont
  {Das}}, \bibinfo {author} {\bibfnamefont {S.}~\bibnamefont {Chakrabarty}},
  \bibinfo {author} {\bibfnamefont {A.}~\bibnamefont {Dhar}}, \bibinfo {author}
  {\bibfnamefont {A.}~\bibnamefont {Kundu}}, \bibinfo {author} {\bibfnamefont
  {D.~A.}\ \bibnamefont {Huse}}, \bibinfo {author} {\bibfnamefont
  {R.}~\bibnamefont {Moessner}}, \bibinfo {author} {\bibfnamefont {S.~S.}\
  \bibnamefont {Ray}}, \ and\ \bibinfo {author} {\bibfnamefont
  {S.}~\bibnamefont {Bhattacharjee}},\ }\href {\doibase
  10.1103/PhysRevLett.121.024101} {\bibfield  {journal} {\bibinfo  {journal}
  {\textcolor{blue}{Phys. Rev. Lett.}}\ }\textbf {\bibinfo {volume} {121}},\
  \bibinfo {pages} {024101} (\bibinfo {year} {2018})}\BibitemShut {NoStop}%
\bibitem [{\citenamefont {Khemani}\ \emph {et~al.}(2018)\citenamefont
  {Khemani}, \citenamefont {Huse},\ and\ \citenamefont
  {Nahum}}]{2018-khemani--nahum}%
  \BibitemOpen
  \bibfield  {author} {\bibinfo {author} {\bibfnamefont {V.}~\bibnamefont
  {Khemani}}, \bibinfo {author} {\bibfnamefont {D.~A.}\ \bibnamefont {Huse}}, \
  and\ \bibinfo {author} {\bibfnamefont {A.}~\bibnamefont {Nahum}},\ }\href
  {\doibase 10.1103/PhysRevB.98.144304} {\bibfield  {journal} {\bibinfo
  {journal} {{Phys. Rev. B}}\ }\textbf {\bibinfo {volume} {98}},\ \bibinfo
  {pages} {144304} (\bibinfo {year} {2018})}\BibitemShut {NoStop}%
\bibitem [{\citenamefont {Bilitewski}\ \emph {et~al.}(2018)\citenamefont
  {Bilitewski}, \citenamefont {Bhattacharjee},\ and\ \citenamefont
  {Moessner}}]{2018-bilitweski--moessner}%
  \BibitemOpen
  \bibfield  {author} {\bibinfo {author} {\bibfnamefont {T.}~\bibnamefont
  {Bilitewski}}, \bibinfo {author} {\bibfnamefont {S.}~\bibnamefont
  {Bhattacharjee}}, \ and\ \bibinfo {author} {\bibfnamefont {R.}~\bibnamefont
  {Moessner}},\ }\href {\doibase 10.1103/PhysRevLett.121.250602} {\bibfield
  {journal} {\bibinfo  {journal} {{Phys. Rev. Lett.}}\ }\textbf {\bibinfo
  {volume} {121}},\ \bibinfo {pages} {250602} (\bibinfo {year}
  {2018})}\BibitemShut {NoStop}%
\bibitem [{\citenamefont {Jalabert}\ \emph {et~al.}(2018)\citenamefont
  {Jalabert}, \citenamefont {Garc\'{\i}a-Mata},\ and\ \citenamefont
  {Wisniacki}}]{2018-jalabert--wisniacki}%
  \BibitemOpen
  \bibfield  {author} {\bibinfo {author} {\bibfnamefont {R.~A.}\ \bibnamefont
  {Jalabert}}, \bibinfo {author} {\bibfnamefont {I.}~\bibnamefont
  {Garc\'{\i}a-Mata}}, \ and\ \bibinfo {author} {\bibfnamefont {D.~A.}\
  \bibnamefont {Wisniacki}},\ }\href {\doibase 10.1103/PhysRevE.98.062218}
  {\bibfield  {journal} {\bibinfo  {journal} {{Phys. Rev. E}}\ }\textbf
  {\bibinfo {volume} {98}},\ \bibinfo {pages} {062218} (\bibinfo {year}
  {2018})}\BibitemShut {NoStop}%
\bibitem [{\citenamefont {Ch\'avez-Carlos}\ \emph {et~al.}(2019)\citenamefont
  {Ch\'avez-Carlos}, \citenamefont {L\'opez-del Carpio}, \citenamefont
  {Bastarrachea-Magnani}, \citenamefont {Str\'ansk\'y}, \citenamefont
  {Lerma-Hern\'andez}, \citenamefont {Santos},\ and\ \citenamefont
  {Hirsch}}]{2019-chavez--hirsch}%
  \BibitemOpen
  \bibfield  {author} {\bibinfo {author} {\bibfnamefont {J.}~\bibnamefont
  {Ch\'avez-Carlos}}, \bibinfo {author} {\bibfnamefont {B.}~\bibnamefont
  {L\'opez-del Carpio}}, \bibinfo {author} {\bibfnamefont {M.~A.}\ \bibnamefont
  {Bastarrachea-Magnani}}, \bibinfo {author} {\bibfnamefont {P.}~\bibnamefont
  {Str\'ansk\'y}}, \bibinfo {author} {\bibfnamefont {S.}~\bibnamefont
  {Lerma-Hern\'andez}}, \bibinfo {author} {\bibfnamefont {L.~F.}\ \bibnamefont
  {Santos}}, \ and\ \bibinfo {author} {\bibfnamefont {J.~G.}\ \bibnamefont
  {Hirsch}},\ }\href {\doibase 10.1103/PhysRevLett.122.024101} {\bibfield
  {journal} {\bibinfo  {journal} {\textcolor{blue}{Phys. Rev. Lett.}}\ }\textbf
  {\bibinfo {volume} {122}},\ \bibinfo {pages} {024101} (\bibinfo {year}
  {2019})}\BibitemShut {NoStop}%
\bibitem [{\citenamefont {Chatterjee}\ \emph {et~al.}(2020)\citenamefont
  {Chatterjee}, \citenamefont {Kundu},\ and\ \citenamefont
  {Kulkarni}}]{2020-chatterjee--kulkarni}%
  \BibitemOpen
  \bibfield  {author} {\bibinfo {author} {\bibfnamefont {A.~K.}\ \bibnamefont
  {Chatterjee}}, \bibinfo {author} {\bibfnamefont {A.}~\bibnamefont {Kundu}}, \
  and\ \bibinfo {author} {\bibfnamefont {M.}~\bibnamefont {Kulkarni}},\ }\href
  {\doibase 10.1103/PhysRevE.102.052103} {\bibfield  {journal} {\bibinfo
  {journal} {\textcolor{blue}{Phys. Rev. E}}\ }\textbf {\bibinfo {volume}
  {102}},\ \bibinfo {pages} {052103} (\bibinfo {year} {2020})}\BibitemShut
  {NoStop}%
\bibitem [{\citenamefont {Kumar}\ \emph {et~al.}(2020)\citenamefont {Kumar},
  \citenamefont {Kundu}, \citenamefont {Kulkarni}, \citenamefont {Huse},\ and\
  \citenamefont {Dhar}}]{2020-kumar--dhar}%
  \BibitemOpen
  \bibfield  {author} {\bibinfo {author} {\bibfnamefont {M.}~\bibnamefont
  {Kumar}}, \bibinfo {author} {\bibfnamefont {A.}~\bibnamefont {Kundu}},
  \bibinfo {author} {\bibfnamefont {M.}~\bibnamefont {Kulkarni}}, \bibinfo
  {author} {\bibfnamefont {D.~A.}\ \bibnamefont {Huse}}, \ and\ \bibinfo
  {author} {\bibfnamefont {A.}~\bibnamefont {Dhar}},\ }\href {\doibase
  10.1103/PhysRevE.102.022130} {\bibfield  {journal} {\bibinfo  {journal}
  {\textcolor{blue}{Phys. Rev. E}}\ }\textbf {\bibinfo {volume} {102}},\
  \bibinfo {pages} {022130} (\bibinfo {year} {2020})}\BibitemShut {NoStop}%
\bibitem [{\citenamefont {Ruidas}\ and\ \citenamefont
  {Banerjee}(2021)}]{2021-ruidas-banerjee}%
  \BibitemOpen
  \bibfield  {author} {\bibinfo {author} {\bibfnamefont {S.}~\bibnamefont
  {Ruidas}}\ and\ \bibinfo {author} {\bibfnamefont {S.}~\bibnamefont
  {Banerjee}},\ }\href {\doibase 10.21468/SciPostPhys.11.5.087} {\bibfield
  {journal} {\bibinfo  {journal} {SciPost Phys.}\ }\textbf {\bibinfo {volume}
  {11}},\ \bibinfo {pages} {87} (\bibinfo {year} {2021})}\BibitemShut {NoStop}%
\bibitem [{\citenamefont {Bilitewski}\ \emph {et~al.}(2021)\citenamefont
  {Bilitewski}, \citenamefont {Bhattacharjee},\ and\ \citenamefont
  {Moessner}}]{2021-bilitewski--moessner}%
  \BibitemOpen
  \bibfield  {author} {\bibinfo {author} {\bibfnamefont {T.}~\bibnamefont
  {Bilitewski}}, \bibinfo {author} {\bibfnamefont {S.}~\bibnamefont
  {Bhattacharjee}}, \ and\ \bibinfo {author} {\bibfnamefont {R.}~\bibnamefont
  {Moessner}},\ }\href {\doibase 10.1103/PhysRevB.103.174302} {\bibfield
  {journal} {\bibinfo  {journal} {\textcolor{blue}{Phys. Rev. B}}\ }\textbf
  {\bibinfo {volume} {103}},\ \bibinfo {pages} {174302} (\bibinfo {year}
  {2021})}\BibitemShut {NoStop}%
\bibitem [{\citenamefont {S.}\ \emph {et~al.}(2021)\citenamefont {S.},
  \citenamefont {Huse},\ and\ \citenamefont {Kulkarni}}]{2021-s--kulkarni}%
  \BibitemOpen
  \bibfield  {author} {\bibinfo {author} {\bibfnamefont {B.~K.}\ \bibnamefont
  {S.}}, \bibinfo {author} {\bibfnamefont {D.~A.}\ \bibnamefont {Huse}}, \ and\
  \bibinfo {author} {\bibfnamefont {M.}~\bibnamefont {Kulkarni}},\ }\href
  {\doibase 10.1103/PhysRevE.104.044117} {\bibfield  {journal} {\bibinfo
  {journal} {Phys. Rev. E}\ }\textbf {\bibinfo {volume} {104}},\ \bibinfo
  {pages} {044117} (\bibinfo {year} {2021})}\BibitemShut {NoStop}%
\bibitem [{\citenamefont {Gerling}\ and\ \citenamefont
  {Landau}(1990)}]{1990-gerling-landau}%
  \BibitemOpen
  \bibfield  {author} {\bibinfo {author} {\bibfnamefont {R.~W.}\ \bibnamefont
  {Gerling}}\ and\ \bibinfo {author} {\bibfnamefont {D.~P.}\ \bibnamefont
  {Landau}},\ }\href {\doibase 10.1103/PhysRevB.42.8214} {\bibfield  {journal}
  {\bibinfo  {journal} {Phys. Rev. B}\ }\textbf {\bibinfo {volume} {42}},\
  \bibinfo {pages} {8214} (\bibinfo {year} {1990})}\BibitemShut {NoStop}%
\bibitem [{\citenamefont {Liu}\ \emph {et~al.}(1991)\citenamefont {Liu},
  \citenamefont {Srivastava}, \citenamefont {Viswanath},\ and\ \citenamefont
  {Müller}}]{1991-liu--muller}%
  \BibitemOpen
  \bibfield  {author} {\bibinfo {author} {\bibfnamefont {J.}~\bibnamefont
  {Liu}}, \bibinfo {author} {\bibfnamefont {N.}~\bibnamefont {Srivastava}},
  \bibinfo {author} {\bibfnamefont {V.~S.}\ \bibnamefont {Viswanath}}, \ and\
  \bibinfo {author} {\bibfnamefont {G.}~\bibnamefont {Müller}},\ }\href
  {\doibase 10.1063/1.350037} {\bibfield  {journal} {\bibinfo  {journal}
  {Journal of Applied Physics}\ }\textbf {\bibinfo {volume} {70}},\ \bibinfo
  {pages} {6181} (\bibinfo {year} {1991})}\BibitemShut {NoStop}%
\bibitem [{\citenamefont {de~Alcantara~Bonfim}\ and\ \citenamefont
  {Reiter}(1992)}]{1992-bonfim-reiter}%
  \BibitemOpen
  \bibfield  {author} {\bibinfo {author} {\bibfnamefont {O.~F.}\ \bibnamefont
  {de~Alcantara~Bonfim}}\ and\ \bibinfo {author} {\bibfnamefont
  {G.}~\bibnamefont {Reiter}},\ }\href {\doibase 10.1103/PhysRevLett.69.367}
  {\bibfield  {journal} {\bibinfo  {journal} {Phys. Rev. Lett.}\ }\textbf
  {\bibinfo {volume} {69}},\ \bibinfo {pages} {367} (\bibinfo {year}
  {1992})}\BibitemShut {NoStop}%
\bibitem [{\citenamefont {B\"ohm}\ \emph {et~al.}(1993)\citenamefont {B\"ohm},
  \citenamefont {Gerling},\ and\ \citenamefont {Leschke}}]{1993-bohm--leschke}%
  \BibitemOpen
  \bibfield  {author} {\bibinfo {author} {\bibfnamefont {M.}~\bibnamefont
  {B\"ohm}}, \bibinfo {author} {\bibfnamefont {R.~W.}\ \bibnamefont {Gerling}},
  \ and\ \bibinfo {author} {\bibfnamefont {H.}~\bibnamefont {Leschke}},\ }\href
  {\doibase 10.1103/PhysRevLett.70.248} {\bibfield  {journal} {\bibinfo
  {journal} {Phys. Rev. Lett.}\ }\textbf {\bibinfo {volume} {70}},\ \bibinfo
  {pages} {248} (\bibinfo {year} {1993})}\BibitemShut {NoStop}%
\bibitem [{\citenamefont {Lovesey}\ and\ \citenamefont
  {Balcar}(1994)}]{1994-lovesey-balcar}%
  \BibitemOpen
  \bibfield  {author} {\bibinfo {author} {\bibfnamefont {S.~W.}\ \bibnamefont
  {Lovesey}}\ and\ \bibinfo {author} {\bibfnamefont {E.}~\bibnamefont
  {Balcar}},\ }\href {\doibase 10.1088/0953-8984/6/6/027} {\bibfield  {journal}
  {\bibinfo  {journal} {Journal of Physics: Condensed Matter}\ }\textbf
  {\bibinfo {volume} {6}},\ \bibinfo {pages} {1253} (\bibinfo {year}
  {1994})}\BibitemShut {NoStop}%
\bibitem [{\citenamefont {Srivastava}\ \emph {et~al.}(1994)\citenamefont
  {Srivastava}, \citenamefont {Liu}, \citenamefont {Viswanath},\ and\
  \citenamefont {Müller}}]{1994-srivastava--muller}%
  \BibitemOpen
  \bibfield  {author} {\bibinfo {author} {\bibfnamefont {N.}~\bibnamefont
  {Srivastava}}, \bibinfo {author} {\bibfnamefont {J.}~\bibnamefont {Liu}},
  \bibinfo {author} {\bibfnamefont {V.~S.}\ \bibnamefont {Viswanath}}, \ and\
  \bibinfo {author} {\bibfnamefont {G.}~\bibnamefont {Müller}},\ }\href
  {\doibase 10.1063/1.356842} {\bibfield  {journal} {\bibinfo  {journal}
  {Journal of Applied Physics}\ }\textbf {\bibinfo {volume} {75}},\ \bibinfo
  {pages} {6751} (\bibinfo {year} {1994})}\BibitemShut {NoStop}%
\bibitem [{\citenamefont {Meister}\ \emph {et~al.}(2001)\citenamefont
  {Meister}, \citenamefont {Mertens},\ and\ \citenamefont
  {S\'{a}nchez}}]{2001-meister--sanchez}%
  \BibitemOpen
  \bibfield  {author} {\bibinfo {author} {\bibfnamefont {M.}~\bibnamefont
  {Meister}}, \bibinfo {author} {\bibfnamefont {F.}~\bibnamefont {Mertens}}, \
  and\ \bibinfo {author} {\bibfnamefont {A.}~\bibnamefont {S\'{a}nchez}},\
  }\href {\doibase 10.1007/s100510170259} {\bibfield  {journal} {\bibinfo
  {journal} {The European Physical Journal B - Condensed Matter and Complex
  Systems}\ }\textbf {\bibinfo {volume} {20}},\ \bibinfo {pages} {405}
  (\bibinfo {year} {2001})}\BibitemShut {NoStop}%
\bibitem [{\citenamefont {Bagchi}(2013)}]{2013-bagchi}%
  \BibitemOpen
  \bibfield  {author} {\bibinfo {author} {\bibfnamefont {D.}~\bibnamefont
  {Bagchi}},\ }\href {\doibase 10.1103/PhysRevB.87.075133} {\bibfield
  {journal} {\bibinfo  {journal} {Phys. Rev. B}\ }\textbf {\bibinfo {volume}
  {87}},\ \bibinfo {pages} {075133} (\bibinfo {year} {2013})}\BibitemShut
  {NoStop}%
\bibitem [{\citenamefont {Li}(2019)}]{2019-li}%
  \BibitemOpen
  \bibfield  {author} {\bibinfo {author} {\bibfnamefont {N.}~\bibnamefont
  {Li}},\ }\href {\doibase 10.1103/PhysRevE.100.062104} {\bibfield  {journal}
  {\bibinfo  {journal} {Phys. Rev. E}\ }\textbf {\bibinfo {volume} {100}},\
  \bibinfo {pages} {062104} (\bibinfo {year} {2019})}\BibitemShut {NoStop}%
\bibitem [{\citenamefont {De~Nardis}\ \emph
  {et~al.}(2020{\natexlab{b}})\citenamefont {De~Nardis}, \citenamefont
  {Medenjak}, \citenamefont {Karrasch},\ and\ \citenamefont
  {Ilievski}}]{2020-nardis--ilievski}%
  \BibitemOpen
  \bibfield  {author} {\bibinfo {author} {\bibfnamefont {J.}~\bibnamefont
  {De~Nardis}}, \bibinfo {author} {\bibfnamefont {M.}~\bibnamefont {Medenjak}},
  \bibinfo {author} {\bibfnamefont {C.}~\bibnamefont {Karrasch}}, \ and\
  \bibinfo {author} {\bibfnamefont {E.}~\bibnamefont {Ilievski}},\ }\href
  {\doibase 10.1103/PhysRevLett.124.210605} {\bibfield  {journal} {\bibinfo
  {journal} {Phys. Rev. Lett.}\ }\textbf {\bibinfo {volume} {124}},\ \bibinfo
  {pages} {210605} (\bibinfo {year} {2020}{\natexlab{b}})}\BibitemShut
  {NoStop}%
\bibitem [{\citenamefont {Glorioso}\ \emph {et~al.}(2021)\citenamefont
  {Glorioso}, \citenamefont {Delacrétaz}, \citenamefont {Chen}, \citenamefont
  {Nandkishore},\ and\ \citenamefont {Lucas}}]{2021-glorioso--lucas}%
  \BibitemOpen
  \bibfield  {author} {\bibinfo {author} {\bibfnamefont {P.}~\bibnamefont
  {Glorioso}}, \bibinfo {author} {\bibfnamefont {L.~V.}\ \bibnamefont
  {Delacrétaz}}, \bibinfo {author} {\bibfnamefont {X.}~\bibnamefont {Chen}},
  \bibinfo {author} {\bibfnamefont {R.~M.}\ \bibnamefont {Nandkishore}}, \ and\
  \bibinfo {author} {\bibfnamefont {A.}~\bibnamefont {Lucas}},\ }\href
  {\doibase 10.21468/SciPostPhys.10.1.015} {\bibfield  {journal} {\bibinfo
  {journal} {SciPost Phys.}\ }\textbf {\bibinfo {volume} {10}},\ \bibinfo
  {pages} {15} (\bibinfo {year} {2021})}\BibitemShut {NoStop}%
\bibitem [{\citenamefont {McRoberts}\ \emph {et~al.}(2022)\citenamefont
  {McRoberts}, \citenamefont {Bilitewski}, \citenamefont {Haque},\ and\
  \citenamefont {Moessner}}]{2022-mcroberts--moessner}%
  \BibitemOpen
  \bibfield  {author} {\bibinfo {author} {\bibfnamefont {A.~J.}\ \bibnamefont
  {McRoberts}}, \bibinfo {author} {\bibfnamefont {T.}~\bibnamefont
  {Bilitewski}}, \bibinfo {author} {\bibfnamefont {M.}~\bibnamefont {Haque}}, \
  and\ \bibinfo {author} {\bibfnamefont {R.}~\bibnamefont {Moessner}},\ }\href
  {\doibase 10.1103/PhysRevB.105.L100403} {\bibfield  {journal} {\bibinfo
  {journal} {Phys. Rev. B}\ }\textbf {\bibinfo {volume} {105}},\ \bibinfo
  {pages} {L100403} (\bibinfo {year} {2022})}\BibitemShut {NoStop}%
\end{thebibliography}%


%merlin.mbs apsrev4-1.bst 2010-07-25 4.21a (PWD, AO, DPC) hacked
%Control: key (0)
%Control: author (8) initials jnrlst
%Control: editor formatted (1) identically to author
%Control: production of article title (-1) disabled
%Control: page (0) single
%Control: year (1) truncated
%Control: production of eprint (0) enabled
\begin{thebibliography}{4}%
\makeatletter
\providecommand \@ifxundefined [1]{%
 \@ifx{#1\undefined}
}%
\providecommand \@ifnum [1]{%
 \ifnum #1\expandafter \@firstoftwo
 \else \expandafter \@secondoftwo
 \fi
}%
\providecommand \@ifx [1]{%
 \ifx #1\expandafter \@firstoftwo
 \else \expandafter \@secondoftwo
 \fi
}%
\providecommand \natexlab [1]{#1}%
\providecommand \enquote  [1]{``#1''}%
\providecommand \bibnamefont  [1]{#1}%
\providecommand \bibfnamefont [1]{#1}%
\providecommand \citenamefont [1]{#1}%
\providecommand \href@noop [0]{\@secondoftwo}%
\providecommand \href [0]{\begingroup \@sanitize@url \@href}%
\providecommand \@href[1]{\@@startlink{#1}\@@href}%
\providecommand \@@href[1]{\endgroup#1\@@endlink}%
\providecommand \@sanitize@url [0]{\catcode `\\12\catcode `\$12\catcode
  `\&12\catcode `\#12\catcode `\^12\catcode `\_12\catcode `\%12\relax}%
\providecommand \@@startlink[1]{}%
\providecommand \@@endlink[0]{}%
\providecommand \url  [0]{\begingroup\@sanitize@url \@url }%
\providecommand \@url [1]{\endgroup\@href {#1}{\urlprefix }}%
\providecommand \urlprefix  [0]{URL }%
\providecommand \Eprint [0]{\href }%
\providecommand \doibase [0]{http://dx.doi.org/}%
\providecommand \selectlanguage [0]{\@gobble}%
\providecommand \bibinfo  [0]{\@secondoftwo}%
\providecommand \bibfield  [0]{\@secondoftwo}%
\providecommand \translation [1]{[#1]}%
\providecommand \BibitemOpen [0]{}%
\providecommand \bibitemStop [0]{}%
\providecommand \bibitemNoStop [0]{.\EOS\space}%
\providecommand \EOS [0]{\spacefactor3000\relax}%
\providecommand \BibitemShut  [1]{\csname bibitem#1\endcsname}%
\let\auto@bib@innerbib\@empty
%</preamble>
\bibitem [{\citenamefont {Alzate-Cardona}\ \emph {et~al.}(2019)\citenamefont
  {Alzate-Cardona}, \citenamefont {Sabogal-Su{\'{a}}rez}, \citenamefont
  {Evans},\ and\ \citenamefont {Restrepo-Parra}}]{2019-cardona--parra}%
  \BibitemOpen
  \bibfield  {author} {\bibinfo {author} {\bibfnamefont {J.~D.}\ \bibnamefont
  {Alzate-Cardona}}, \bibinfo {author} {\bibfnamefont {D.}~\bibnamefont
  {Sabogal-Su{\'{a}}rez}}, \bibinfo {author} {\bibfnamefont {R.~F.~L.}\
  \bibnamefont {Evans}}, \ and\ \bibinfo {author} {\bibfnamefont
  {E.}~\bibnamefont {Restrepo-Parra}},\ }\href {\doibase
  10.1088/1361-648x/aaf852} {\bibfield  {journal} {\bibinfo  {journal} {Journal
  of Physics: Condensed Matter}\ }\textbf {\bibinfo {volume} {31}},\ \bibinfo
  {pages} {095802} (\bibinfo {year} {2019})}\BibitemShut {NoStop}%
\bibitem [{\citenamefont {Das}\ \emph {et~al.}(2019)\citenamefont {Das},
  \citenamefont {Kulkarni}, \citenamefont {Spohn},\ and\ \citenamefont
  {Dhar}}]{2019-das--dhar}%
  \BibitemOpen
  \bibfield  {author} {\bibinfo {author} {\bibfnamefont {A.}~\bibnamefont
  {Das}}, \bibinfo {author} {\bibfnamefont {M.}~\bibnamefont {Kulkarni}},
  \bibinfo {author} {\bibfnamefont {H.}~\bibnamefont {Spohn}}, \ and\ \bibinfo
  {author} {\bibfnamefont {A.}~\bibnamefont {Dhar}},\ }\href {\doibase
  10.1103/PhysRevE.100.042116} {\bibfield  {journal} {\bibinfo  {journal}
  {Phys. Rev. E}\ }\textbf {\bibinfo {volume} {100}},\ \bibinfo {pages}
  {042116} (\bibinfo {year} {2019})}\BibitemShut {NoStop}%
\bibitem [{\citenamefont {Hairer}\ \emph {et~al.}(1993)\citenamefont {Hairer},
  \citenamefont {N\o{}rsett},\ and\ \citenamefont
  {Wanner}}]{1993-hairer--wanner}%
  \BibitemOpen
  \bibfield  {author} {\bibinfo {author} {\bibfnamefont {E.}~\bibnamefont
  {Hairer}}, \bibinfo {author} {\bibfnamefont {S.~P.}\ \bibnamefont
  {N\o{}rsett}}, \ and\ \bibinfo {author} {\bibfnamefont {G.}~\bibnamefont
  {Wanner}},\ }\href {https://link.springer.com/book/10.1007/978-3-540-78862-1}
  {\emph {\bibinfo {title} {{Solving Ordinary Differential Equations I}}}},\
  \bibinfo {edition} {2nd}\ ed.\ (\bibinfo  {publisher} {Springer, Berlin,
  Heidelberg},\ \bibinfo {year} {1993})\BibitemShut {NoStop}%
\bibitem [{\citenamefont {Press}\ \emph {et~al.}(2007)\citenamefont {Press},
  \citenamefont {Teukolsky}, \citenamefont {Vetterling},\ and\ \citenamefont
  {Flannery}}]{2007-press--flannery}%
  \BibitemOpen
  \bibfield  {author} {\bibinfo {author} {\bibfnamefont {W.~H.}\ \bibnamefont
  {Press}}, \bibinfo {author} {\bibfnamefont {S.~A.}\ \bibnamefont
  {Teukolsky}}, \bibinfo {author} {\bibfnamefont {W.~T.}\ \bibnamefont
  {Vetterling}}, \ and\ \bibinfo {author} {\bibfnamefont {B.~P.}\ \bibnamefont
  {Flannery}},\ }\href {https://dl.acm.org/doi/book/10.5555/1403886} {\emph
  {\bibinfo {title} {{Numerical Recipes 3rd Edition: The Art of Scientific
  Computing}}}},\ \bibinfo {edition} {3rd}\ ed.\ (\bibinfo  {publisher}
  {Cambridge University Press},\ \bibinfo {address} {USA},\ \bibinfo {year}
  {2007})\BibitemShut {NoStop}%
\end{thebibliography}%

\end{document}

\newcommand{\titlename}{\underline{\textsc{Supplemental material}}\\ \bigskip Robustness of Kardar-Parisi-Zhang scaling in a classical integrable spin chain with broken integrability}

\title{\titlename}

\author{Dipankar Roy}
\email{dipankar.roy@icts.res.in}
\affiliation{International Centre for Theoretical Sciences, Tata Institute of Fundamental Research, Bangalore 560089, India}
\author{Abhishek Dhar}
\email{abhishek.dhar@icts.res.in}
\affiliation{International Centre for Theoretical Sciences, Tata Institute of Fundamental Research, Bangalore 560089, India}
\author{Herbert Spohn}
\email{spohn@ma.tum.de}
\affiliation{Zentrum Mathematik and Physik Department, Technische Universität München, Garching 85748, Germany}
\author{Manas Kulkarni}
\email{manas.kulkarni@icts.res.in}
\affiliation{International Centre for Theoretical Sciences, Tata Institute of Fundamental Research, Bangalore 560089, India}
\date{\today}

\maketitle

\tableofcontents 

\vspace{0.8cm}

\section{Spin-symmetry preserving next nearest neighbour interaction}

\begin{figure}[htbp!]
	\vspace*{0.cm}
	\begin{center}
		\begin{subfigure}{1\linewidth}
			\includegraphics[width=1.0\linewidth]{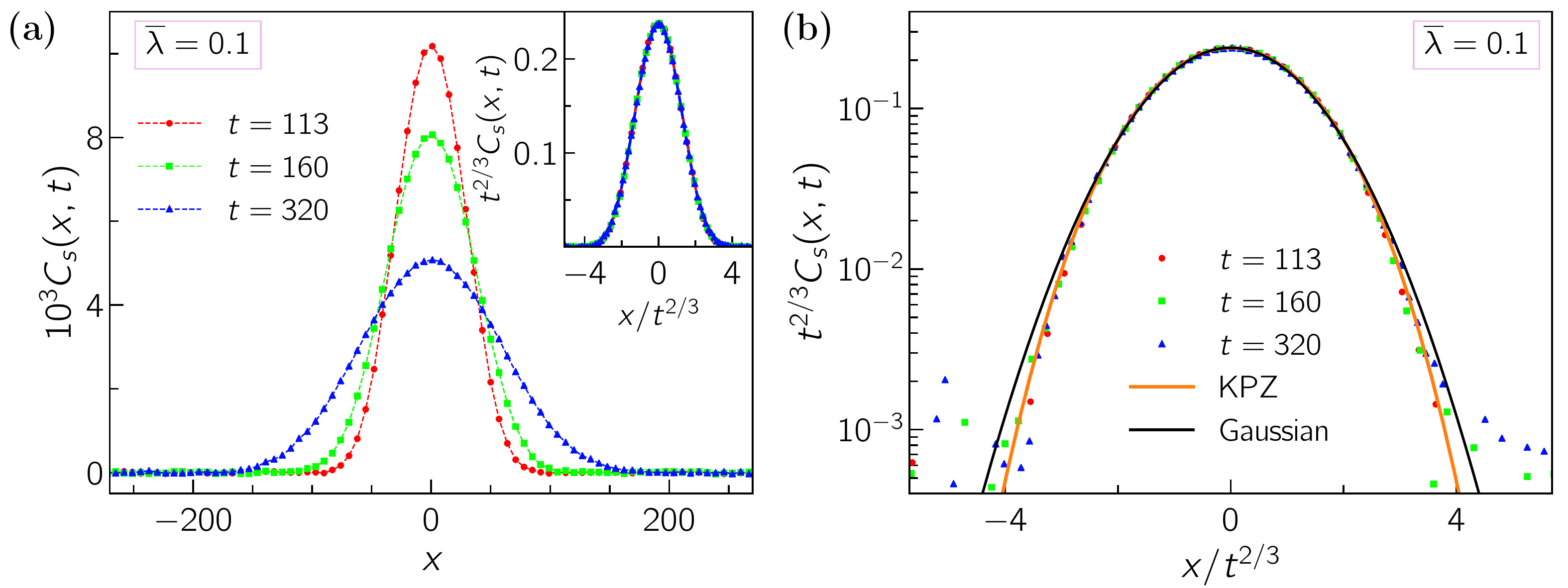}
		\end{subfigure}
		\begin{subfigure}{1\linewidth}
			\includegraphics[width=1.0\linewidth]{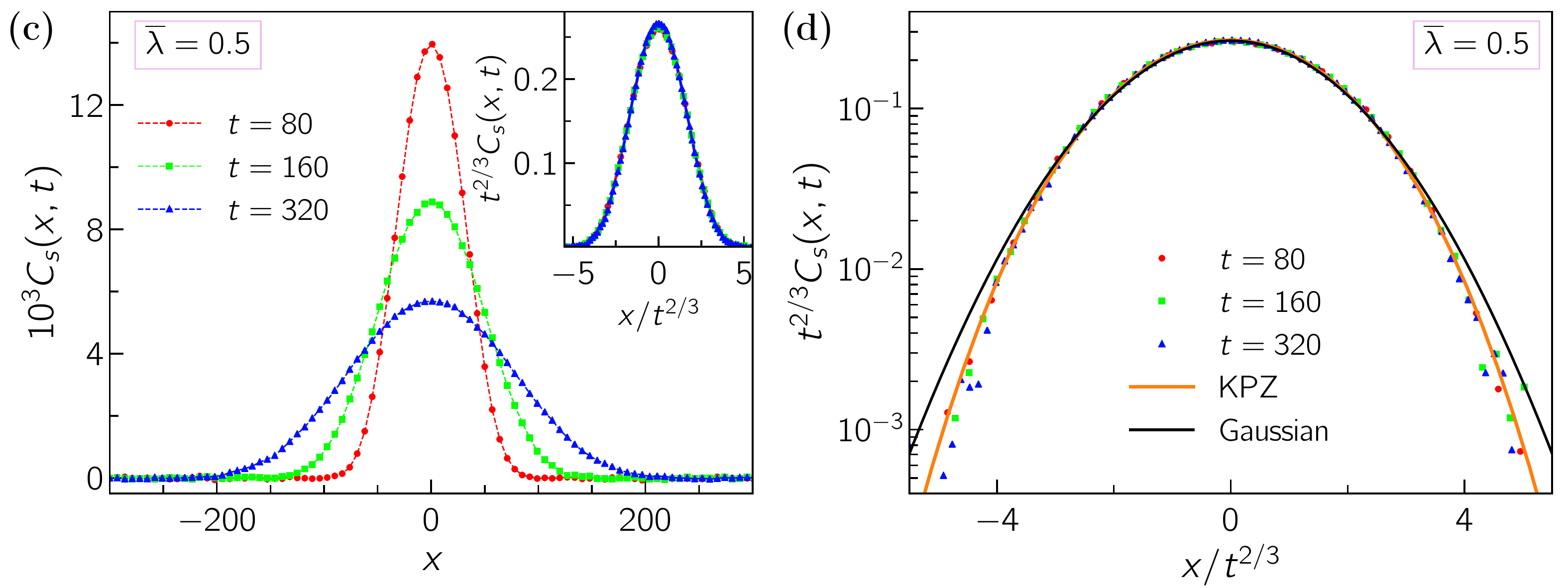}
		\end{subfigure}
		\begin{subfigure}{1\linewidth}
			\includegraphics[width=1\linewidth]{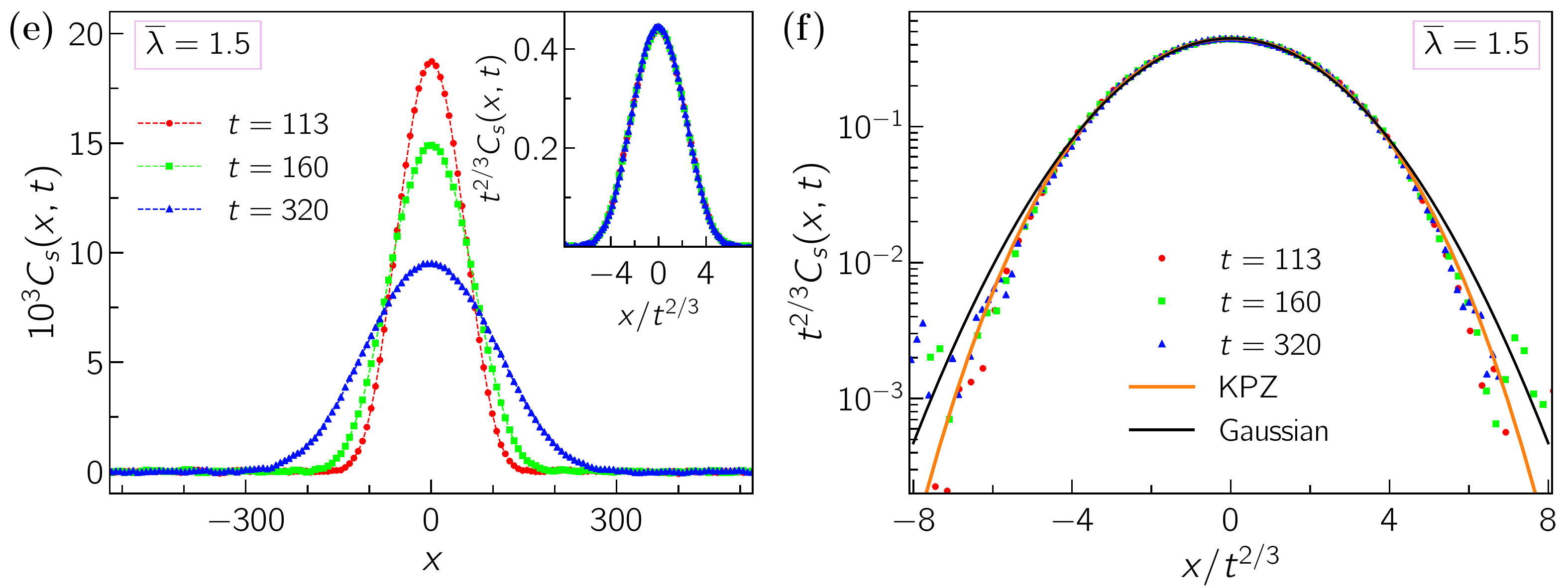}
		\end{subfigure}
	\end{center}
	\caption{(Color online) Plots of the spin correlations for different values of $\overline{\lambda}$ in the case when integrability-breaking perturbation involves next nearest neighbour interaction. We plot $C_{s}(x,t)$ versus $x$ in (a), (c), and (e) and $t^{2/3}C_{s}(x,t)$ versus $x/t^{2/3}$ in (b), (d), and (f). Insets in (a), (c), and (e) show the data collapse on a normal scale. Total number of independent realizations is $10^5$ and $N=2048$.}
	\label{fig:3n-spin}
\end{figure}

\begin{figure}[htbp!]
	\vspace*{0.cm}
	\begin{center}
		\includegraphics[width=\linewidth]{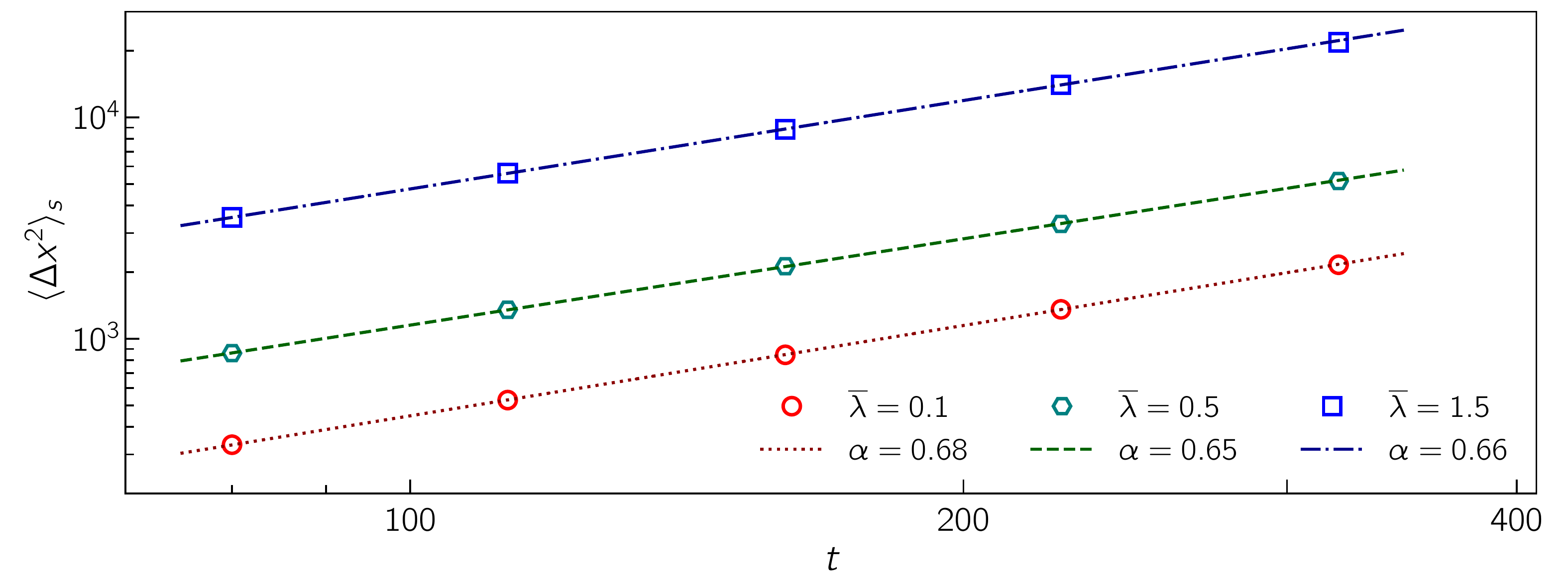}
	\end{center}
	\caption{(Color online) Plots of MSDs (corresponding to the spin correlations in Fig.~\ref{fig:3n-spin}) in the case of next nearest neighbour perturbation for $\overline{\lambda}=0.1, 0.5, 1.5$. We observe that the linear fits are parallel with slopes in agreement with the KPZ exponent.}
	\label{fig:msd-3n}
\end{figure}

In the main text, we study the transport properties in a classical spin chain where spin-symmetry preserving perturbation involving \emph{nearest neighbour} interaction breaks the integrability. We consider here a different class of integrability-breaking, but spin-symmetry preserving perturbation to demonstrate the robustness of KPZ superdiffusion. In particular, we use \emph{next nearest neighbour} interaction to break the integrability by adding a perturbation $\overline{H}_{NI}$ to the Hamiltonian for the ILLL spin chain at the isotropic point:
\begin{equation}
	H =  \sum_{n=1}^{N} - J \ln \left( 1 + \vec{S}_{n} \cdot  \vec{S}_{n+1} \right) + \overline{H}_{NI} %\sum_{n=1}^{N} h( \vec{S}_n , \vec{S}_{n+1}),
	\label{eq:ham2}
\end{equation} 
where $\overline{H}_{NI}$ is given by
\begin{equation}
	\overline{H}_{NI} = - \overline{\lambda} \sum_{n=1}^{N} \vec{S}_{n} \cdot \vec{S}_{n+2}, \quad \overline{\lambda} \in \mathbb{R} . 
\end{equation}
Here, $\overline{\lambda}$ is the strength of perturbation. We observe KPZ behaviour for spin correlation in this case too. The spin correlation and the pertinent scaling is shown in Figs.~\ref{fig:3n-spin}~(a)-(f) for $\overline{\lambda}=0.1,0.5,1.5$. The corresponding MSDs are shown in Fig.~\ref{fig:msd-3n}. We observe exponents close to $2/3$ in these plots for MSDs. These results are for $\beta=1$ (fixed throughout the present work) and should hold true for all temperatures.

Thus, these results along with those in the main text provide compelling evidence for two nontrivial classes of integrability-breaking perturbations in support of the robustness of KPZ superdiffusion when integrability-breaking preserves the symmetry of spins. This leads us to believe that KPZ superdiffusion, observed in the integrable spin chains, survives at least for a large class of symmetry-preserving perturbations which break integrability. However, KPZ superdiffusion is not observed when the symmetry of the spins are not preserved, as we exemplify in the next section.

\section{Breaking integrability without respecting spin-symmetry}

Spin-symmetry plays an essential role in the existence of KPZ superdiffusion in spin chains. In this section, we provide numerical evidence that KPZ superdiffusion is destroyed in the spin chains where integrability is broken with anisotropic perturbations. First we consider a perturbation with \emph{XXZ} interaction, and then with \emph{XYZ}-type anisotropy.

\subsection{\emph{XXZ} integrability-breaking term}

\begin{figure}[htbp!]
	\vspace*{0.cm}
	\begin{center}
		\begin{subfigure}{1\linewidth}
			\includegraphics[width=\linewidth]{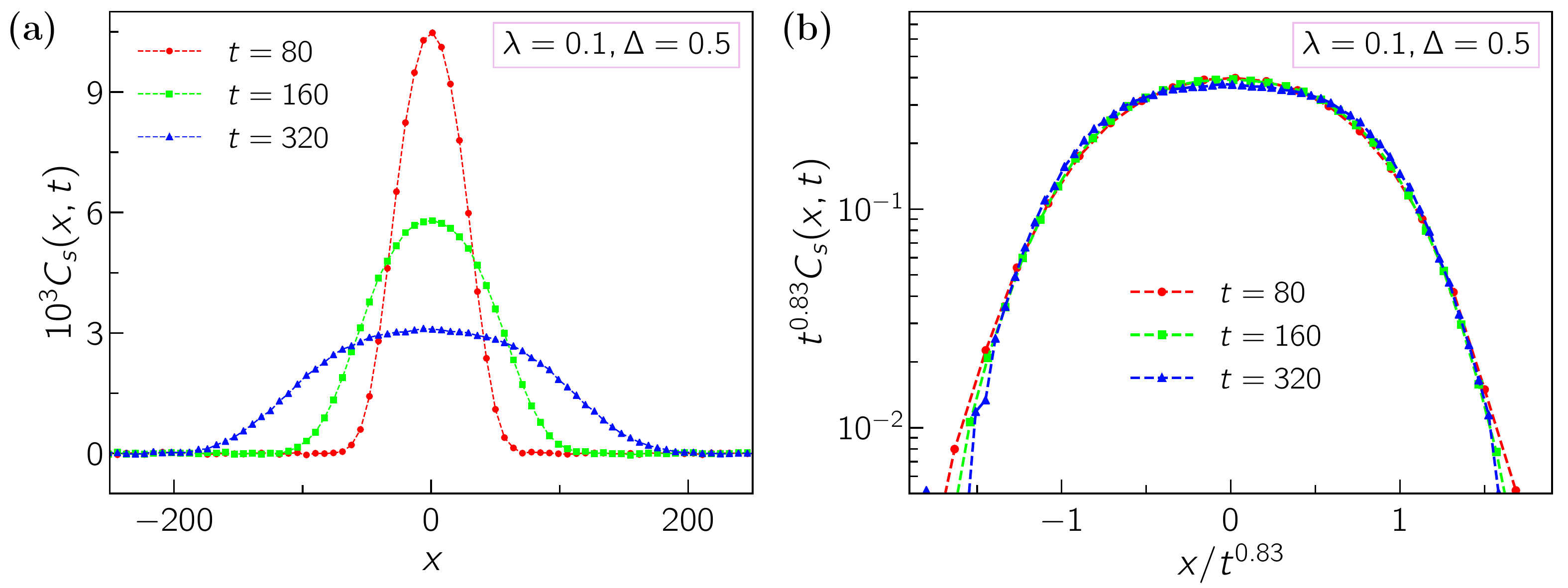}
		\end{subfigure}
		\begin{subfigure}{1\linewidth}
			\includegraphics[width=\linewidth]{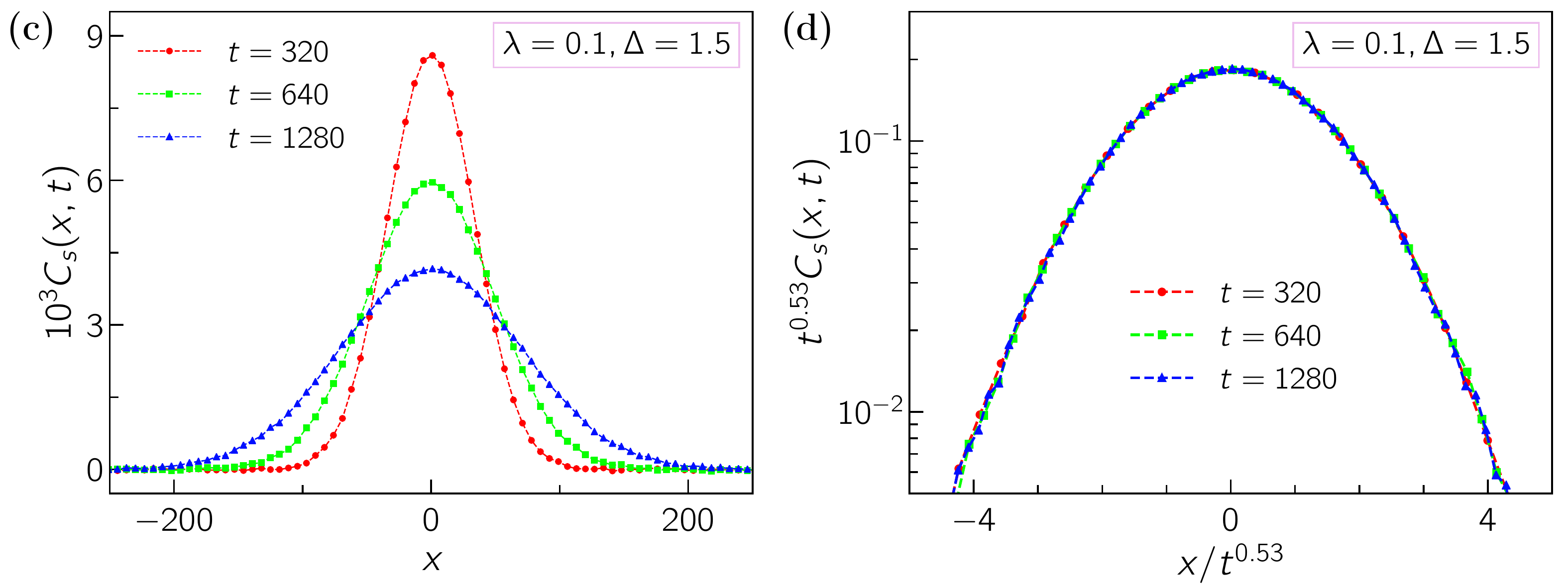}
		\end{subfigure}
	\end{center}
	\caption{(Color online) Plots of the spin correlations in the case of anisotropic integrability-breaking perturbation for different values of $\Delta$ with $\lambda=0.1$. We plot $C_{s}(x,t)$ versus $x$ in (a) and (c).
	In (b), we plot $t^{0.83}C_{s}(x,t)$ versus $x/t^{0.83}$, and we show $t^{0.53}C_{s}(x,t)$ versus $x/t^{0.53}$ in (d).
	Total number of independent realizations is $10^5$ and $N=2048$. This shows immediate deviation from KPZ behaviour when spin-symmetry is broken.}
	\label{fig:ani-spin}
\end{figure}

\begin{figure}[htbp!]
	\begin{center}
		\begin{subfigure}{0.5\linewidth}
			\includegraphics[width=\linewidth]{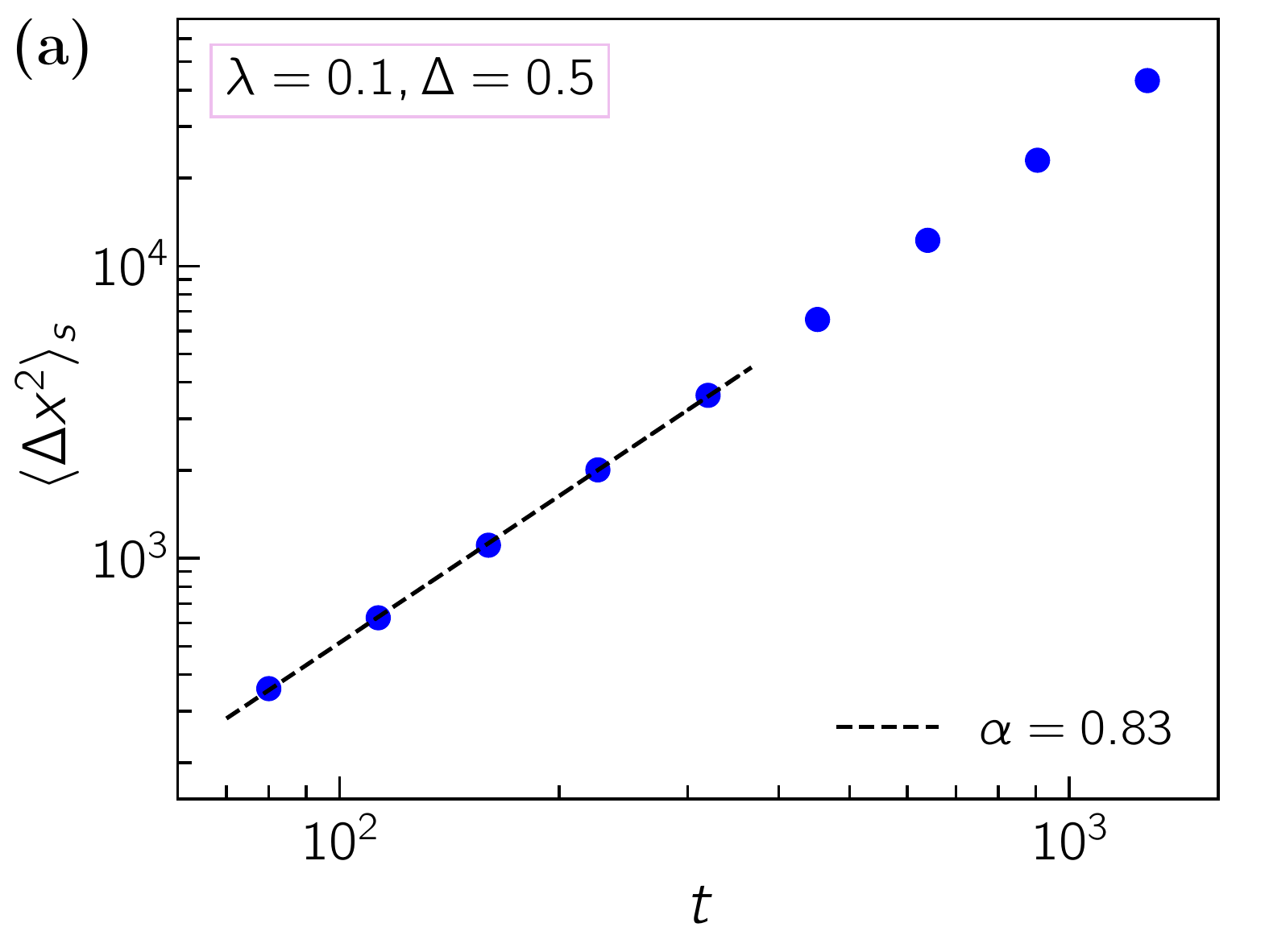}
		\end{subfigure}%
		\begin{subfigure}{0.5\linewidth}
			\includegraphics[width=\linewidth]{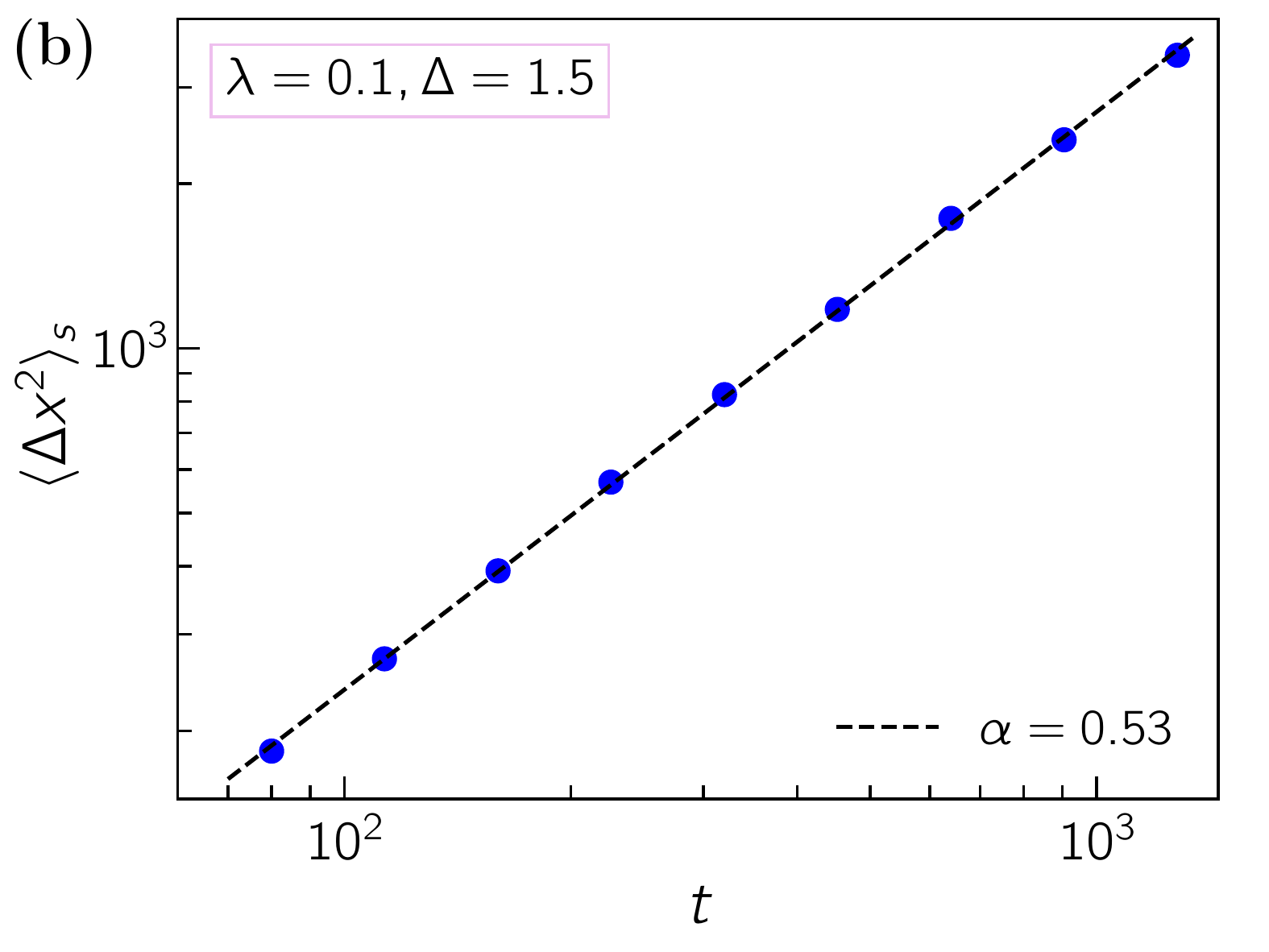}
		\end{subfigure}
	\end{center}
	\caption{(Color online) Plots of MSDs for the spin correlations for anisotropic integrability-breaking perturbation for (a) $\Delta=0.5$ and (b) $\Delta=1.5$.}
	\label{fig:ani-msd}
\end{figure}

We consider a simple \emph{anisotropic} perturbation of \emph{XXZ}-type
\begin{equation}
	H_{NI} =  - \lambda \sum_{n=1}^{N} \left( S_n^x  S_{n+1}^x + S_n^y  S_{n+1}^y + \Delta S_n^z  S_{n+1}^z\right),  
\end{equation}
with $\ \lambda, \Delta \in \mathbb{R}$. Here, $\Delta$ is the anisotropy parameter. If $\Delta =1$, the Hamiltonian is isotropic, whereas the spin chain is anisotropic for $\Delta \neq 1$. Again, $\lambda$ is the strength of the perturbation. We fix the parameters $J=1, \lambda = 0.1,$ and consider the values $\Delta=0.5$ and $\Delta=1.5$ to show numerically that the spin correlation do not show KPZ superdiffusion. We observe a scaling exponent of $\alpha=0.83$ for $\Delta=0.5$ and $\alpha=0.53$ for $\Delta=1.5$ for the spin correlation (see Fig.~\ref{fig:ani-spin}). These exponents are recovered from corresponding MSDs as well (Fig.~\ref{fig:ani-msd}).

\subsection{\emph{XYZ} integrability-breaking term}

\begin{figure}[htbp!]
	\vspace*{0.cm}
	\begin{center}
		\begin{subfigure}{1\linewidth}
			\includegraphics[width=\linewidth]{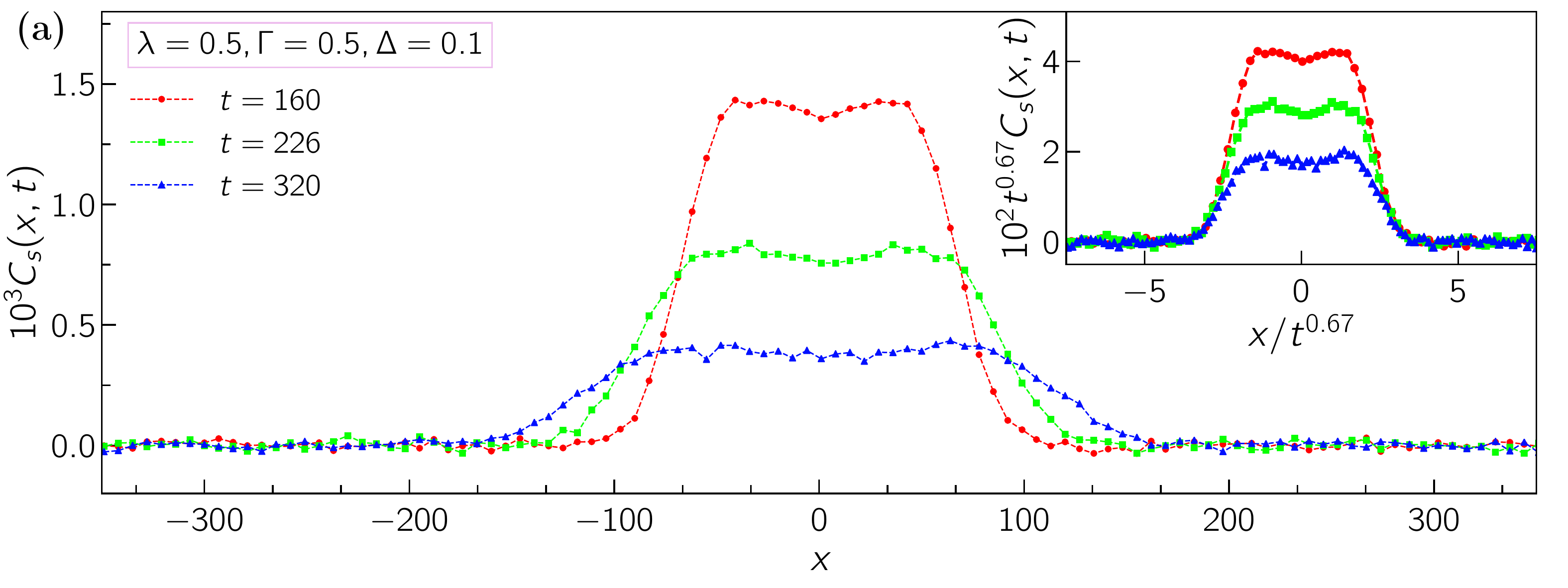}
		\end{subfigure}
		\begin{subfigure}{1\linewidth}
			\includegraphics[width=\linewidth]{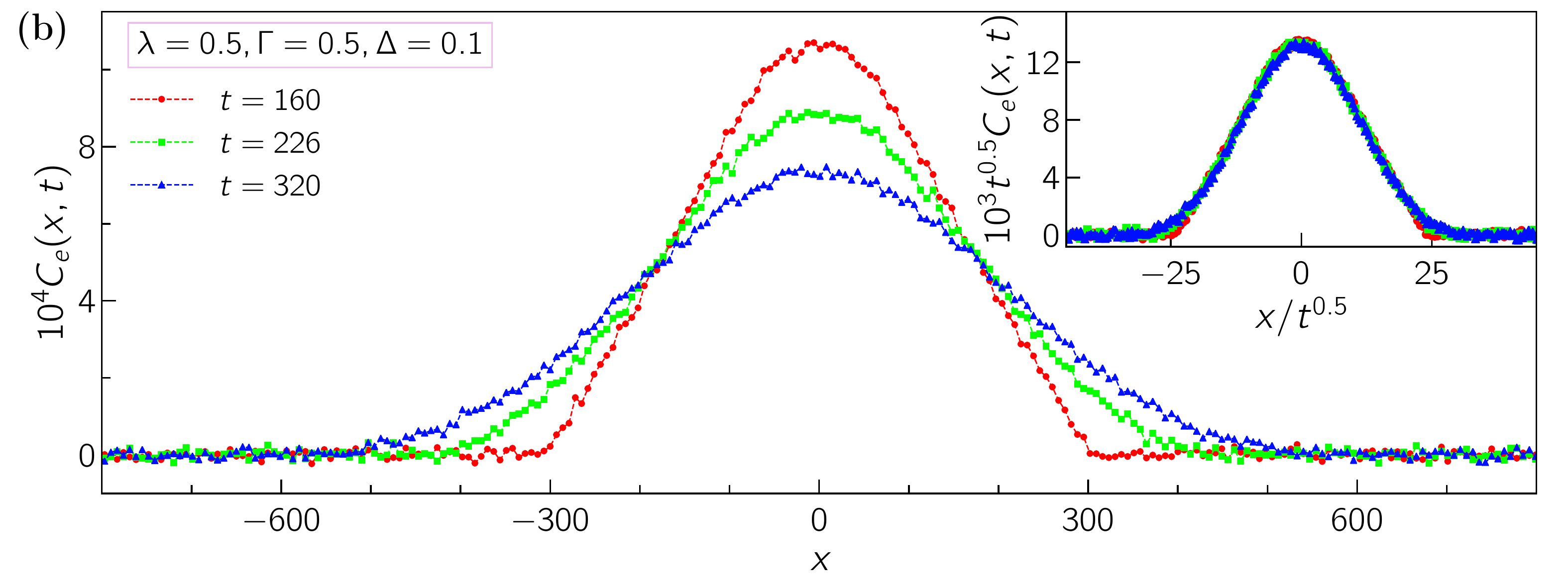}
		\end{subfigure}
	\end{center}
	\caption{(Color online) Plots of (a) spin correlation and (b) energy correlation in the case of \emph{XYZ} anisotropic integrability-breaking perturbation with $\lambda=0.5, \gamma=0.5, \Delta=0.1$. In the inset in (a), we show that spin correlations do not collapse. In fact, spin has no scaling collapse at all due to lack of conservation law. The data collapse for energy correlation in the inset in (b) indicates diffusive behaviour. Total number of independent realizations is $10^5$ and $N=2048$.
	}
	\label{fig:xyz-spin}
\end{figure}

In this section, we consider the case when even total spin component in $z$-direction is  not conserved thereby leaving energy as the only conserved quantity. The integrability-breaking term we consider is given by
\begin{equation}
	H_{NI} =  - \lambda \sum_{n=1}^{N} \left( S_n^x  S_{n+1}^x + \Gamma S_n^y  S_{n+1}^y + \Delta S_n^z  S_{n+1}^z\right), 
	\label{eq:xyz}
\end{equation}
with $\ \lambda, \Gamma, \Delta \in \mathbb{R}$. In particular, we are interested here in the case $\Gamma \neq \Delta \neq 1$. We show the spin correlation in the top panel of Fig.~\ref{fig:xyz-spin}. As expected, the spin correlation functions for different times do not collapse since the quantity $\sum_{x}C_{s}(x,t)$ is not conserved. In the bottom panel of Fig.~\ref{fig:xyz-spin}, we plot energy correlation function whose collapse profile (see inset in Fig.~\ref{fig:xyz-spin}) indicates diffusive behaviour.

\section{Simulation methods \label{sec:simdet}}

We generate the initial conditions for the spin chain at the desired temperature using Metropolis Monte Carlo algorithm. For all the results reported here, we fix $\beta=1$. We use the \emph{adaptive Gaussian move} studied in Ref.~\onlinecite{2019-cardona--parra} to update spins sequentially from left to right. Such a sequence of $N$ updates starting from the leftmost site is referred to as a \emph{Monte Carlo step} (MCS) or a \emph{swipe} \cite{2019-das--dhar}. In the Monte Carlo simulations, we start with an ordered state where all spins point to the positive $z$-axis. Then we perform 5000 initial swipes so that the spin chain reaches thermal equilibrium. After reaching thermal equilibrium, we save the configuration of the spin chain after each 500 successive swipes to ensure the initial conditions are sufficiently decorrelated. These configurations are then used as initial conditions for the dynamical evolution of the spin chain. For the dynamical evolution of the spin chain we use an adaptive Runge-Kutta scheme \cite{1993-hairer--wanner, 2007-press--flannery}. Finally, we average over a large number of initial conditions to compute the spin and energy correlations.

\medskip

\section{Linearization of Hamilton's equations for computing Lyapunov exponents}
We consider linearized equation for a perturbation in the \textit{ip}ILLL spin chain in order to compute the Lyapunov exponents. Let us introduce the notation:
\begin{equation}
    \delta \vec{S}_n = \vec{S}_n^{B}  -  \vec{S}_n^{A} , \quad \epsilon \rightarrow 0,
\end{equation}
where (a) $\vec{S}_n^{A}$ is the original copy evolving from a thermal configuration, and (b) $\vec{S}_n^{B}$ is the perturbed copy evolved from an initial condition obtained after adding perturbation to the initial condition of the original copy at site $N/2$. The perturbation is described in the main text. We derive equations retaining only the terms linear in $\delta \vec{S}_n$ and ignoring higher order terms. In order to write the expressions succinctly, we also introduce the following notations 
\begin{equation}
    \begin{aligned}
    \delta \brac{ \vec{S}_n \times \vec{S}_{n+1} } 
       & \equiv
       \delta \vec{S}_n \times \vec{S}_{n+1} +  \vec{S}_n \times \delta \vec{S}_{n+1} 
       \\ 
       \delta \brac{ \vec{S}_n \cdot \vec{S}_{n+1} } 
       & \equiv
       \delta \vec{S}_n \cdot \vec{S}_{n+1} +  \vec{S}_n \cdot \delta \vec{S}_{n+1} .
    \end{aligned}
\end{equation}
The Hamilton's equations are
\begin{equation}
\begin{aligned}
	\frac{ \text{d} \vec{S}_n }{ \text{d}  t } 
	 & = 
	\vec{S}_n \times
	\Big( 
	\frac{ J \vec{S}_{n-1}}{1 + \vec{S}_{n-1} \cdot \vec{S}_{n} }
	+
	\frac{ J \vec{S}_{n+1}}{1 + \vec{S}_{n} \cdot \vec{S}_{n+1} } 
    \\
	& \ \
	+ 
	\lambda(\vec{S}_{n-1} + \vec{S}_{n+1})
	\Big).
	\label{eq:hamiplll}
\end{aligned}
\end{equation}
Then the linearized equation for $\delta \vec{S}_n$ is 
\begin{equation}
\begin{aligned}
    \frac{ \text{d} \delta\vec{S}_n }{ \text{d}  t } 
	 & = 
	\delta \brac{\vec{S}_n \times \vec{S}_{n-1} }
	\Big( 
	\frac{ J }{1 + \vec{S}_{n-1} \cdot \vec{S}_{n} }
	+ \lambda \Big) 
	\\
	& 
	- \frac{J \vec{S}_n \times \vec{S}_{n-1} }{ (1 + \vec{S}_{n-1} \cdot \vec{S}_{n})^2} \delta \brac{\vec{S}_{n-1} \cdot \vec{S}_{n}}
	\\
	& \ \
	+ \delta \brac{\vec{S}_n \times \vec{S}_{n+1} }
	\Big( 
	\frac{ J }{1 + \vec{S}_{n} \cdot \vec{S}_{n+1} }
	+ \lambda \Big) 
    \\
	& \ \
	- \frac{J \vec{S}_n \times \vec{S}_{n+1} }{ (1 + \vec{S}_{n} \cdot \vec{S}_{n+1})^2} \delta \brac{\vec{S}_{n} \cdot \vec{S}_{n+1}}.
\end{aligned}
\end{equation}
In our simulations, we evolve the above equation simultaneously along with Hamilton's equation in Eq.~\eqref{eq:hamiplll}.

\bibliography{refs.bib}